\documentclass[10pt,twocolumn,letterpaper]{article}

\usepackage{cvpr}              

\usepackage[dvipsnames]{xcolor}

\usepackage{graphicx}
\usepackage{booktabs}
\usepackage{amssymb}
\usepackage{multirow}
\usepackage{makecell}
\usepackage{float}
\usepackage[accsupp]{axessibility}
\definecolor{cvprblue}{rgb}{0.21,0.49,0.74}
\usepackage[pagebackref,breaklinks,colorlinks,citecolor=cvprblue]{hyperref}
\usepackage{listings}
\usepackage[accsupp]{axessibility} 

\def\paperID{*****} 
\def\confName{CVPR}
\def\confYear{2024}

\title{NTIRE 2025 the 2nd Restore Any Image Model (RAIM) in the Wild Challenge}

\author{Jie Liang \and Radu Timofte \and Qiaosi Yi \and Zhengqiang Zhang \and Shuaizheng Liu \and Lingchen Sun \and Rongyuan Wu \and Xindong Zhang \and Hui Zeng \and Lei Zhang \and Tianyu Hao \and Lin Wang \and Zhe Xiao \and Pengzhou Ji \and Shu-Peng Zhong \and Xiangming Wang \and Xiangming Wang \and Jiaqi Yan \and Sishun Pan \and Ce Wang \and Yibin Huang \and ZhanSheng Wang \and Haobo Liang \and Zhenghao Pan \and Jinjian Wu \and Yushen Zuo \and Yuanbo Zhou}

\begin{document}
\maketitle
\begin{abstract}


In this paper, we present a comprehensive overview of the NTIRE 2025 challenge on the 2nd Restore Any Image Model (RAIM) in the Wild. This challenge established a new benchmark for real-world image restoration, featuring diverse scenarios with and without reference ground truth. Participants were tasked with restoring real-captured images suffering from complex and unknown degradations, where both perceptual quality and fidelity were critically evaluated.
The challenge comprised two tracks: (1) the low-light joint denoising and demosaicing (JDD) task, and (2) the image detail enhancement/generation task. Each track included two sub-tasks. The first sub-task involved paired data with available ground truth, enabling quantitative evaluation. The second sub-task dealt with real-world yet unpaired images, emphasizing restoration efficiency and subjective quality assessed through a comprehensive user study.
In total, the challenge attracted nearly 300 registrations, with 51 teams submitting more than 600 results. The top-performing methods advanced the state of the art in image restoration and received unanimous recognition from all 20+ expert judges. The datasets used in Track 1 and Track 2 are available at \url{https://drive.google.com/drive/folders/1Mgqve-yNcE26IIieI8lMIf-25VvZRs_J} and \url{https://drive.google.com/drive/folders/1UB7nnzLwqDZOwDmD9aT8J0KVg2ag4Qae}, respectively. The official challenge pages for Track 1 and Track 2 can be found at \url{https://codalab.lisn.upsaclay.fr/competitions/21334#learn_the_details} and \url{https://codalab.lisn.upsaclay.fr/competitions/21623#learn_the_details}.
\end{abstract}    
\section{Introduction}
\label{sec:intro}

Image restoration, which aims to recover high-quality images from their degraded counterparts, is a fundamental and widely studied task in low-level computer vision. Despite significant progress in recent years, a substantial gap remains between academic research and real-world industrial applications. On the one hand, the image signal processing (ISP) systems in recent high-end digital cameras often encounter complex yet moderate degradations, whereas most academic methods are developed and evaluated under simplified and unrealistic degradation settings. On the other hand, efficiency is critical for deployment on resource-constrained devices such as smartphones, where algorithms must process high-resolution images with limited computational capacity. Consequently, designing and training models that are both efficient and generalizable to real-world scenarios remains a challenging, yet highly valuable, research direction.

Deep learning techniques have substantially advanced the field of image restoration, enabling models to recover high-quality images from their degraded counterparts with remarkable performance. In particular, generative adversarial networks (GANs) have demonstrated strong capabilities in modeling the distribution of natural images, leading to visually realistic and perceptually convincing restoration results. More recently, the emergence of large-scale pre-trained diffusion models has brought a new paradigm to image restoration. These models offer powerful generative priors that can be effectively leveraged to guide the restoration process, producing outputs with improved visual quality, semantic consistency, and fine-grained details. 

In the 1st Restore Any Image Model (RAIM) in the Wild challenge, we provided a platform for researchers to investigate how to bridge the gap between academic research and industrial application. In that challenge, we focused on the real-world image detail enhancement/generation task without any restrictions on efficiency. More than 200 teams participated in the 1st RAIM, and 6 of them won prizes. This year, the Y-Lab of The OPPO Research Institute and the Visual Computing Lab of The Hong Kong Polytechnic University continue to co-host the 2nd RAIM challenge. In this challenge, we will still provide comprehensive data collected in real-world digital photography for researchers to test their models, as well as high-quality feedback from experienced practitioners in the industry. Besides, to better reflect the real-world application, we consider the model efficiency in this challenge.

This challenge aims to provide a platform for both academic and industrial participants to develop, test, and evaluate their algorithms on real-world imaging scenarios, with the goal of bridging the gap between academic research and practical photography applications. The main objectives of the RAIM challenge are as follows:

\begin{itemize}
    \item Construct a benchmark for image restoration in the wild, featuring real-world images with and without reference ground truth across diverse scenarios, along with both objective and subjective evaluation methods.
    \item Promote the research of RAIMs with strong generalization capabilities for handling real-world images.
\end{itemize}


This challenge is one of the NTIRE 2025~\footnote{\url{https://www.cvlai.net/ntire/2025/}} Workshop associated challenges on: ambient lighting normalization~\cite{ntire2025ambient}, reflection removal in the wild~\cite{ntire2025reflection}, shadow removal~\cite{ntire2025shadow}, event-based image deblurring~\cite{ntire2025event}, image denoising~\cite{ntire2025denoising}, XGC quality assessment~\cite{ntire2025xgc}, UGC video enhancement~\cite{ntire2025ugc}, night photography rendering~\cite{ntire2025night}, image super-resolution (x4)~\cite{ntire2025srx4}, real-world face restoration~\cite{ntire2025face}, efficient super-resolution~\cite{ntire2025esr}, HR depth estimation~\cite{ntire2025hrdepth}, efficient burst HDR and restoration~\cite{ntire2025ebhdr}, cross-domain few-shot object detection~\cite{ntire2025cross}, short-form UGC video quality assessment and enhancement~\cite{ntire2025shortugc,ntire2025shortugc_data}, text to image generation model quality assessment~\cite{ntire2025text}, day and night raindrop removal for dual-focused images~\cite{ntire2025day}, video quality assessment for video conferencing~\cite{ntire2025vqe}, low light image enhancement~\cite{ntire2025lowlight}, light field super-resolution~\cite{ntire2025lightfield}, restore any image model (RAIM) in the wild~\cite{ntire2025raim}, raw restoration and super-resolution~\cite{ntire2025raw} and raw reconstruction from RGB on smartphones~\cite{ntire2025rawrgb}.

\renewcommand{\thefootnote}{}
\footnotetext{Jie Liang, Radu Timofte, Qiaosi Yi, Zhengqiang Zhang, Shuaizheng Liu, Lingchen Sun, Rongyuan Wu, Xindong Zhang, Hui Zeng, and Lei Zhang are the organizers of the NTIRE 2025 challenge, and other authors are the participants.}
\footnotetext{The Appendix lists the authors’ teams and affiliations.}

\section{NTIRE 2025 the 2nd RAIM Challenge}
\label{sec:challenge}

\subsection{Training Data}
In this challenge, participants can train their models using any data they can collect and any pre-trained models they can access, as in the 1st RAIM.

\subsection{Validation and Test Data}

To facilitate the design and development of RAIM by participants, we provide two types of validation and test data: paired data with reference ground-truth (R-GT), and unpaired data.

\subsubsection{Paired Data with R-GT}

\begin{figure*}
  \centering
  \includegraphics[width=1\textwidth]{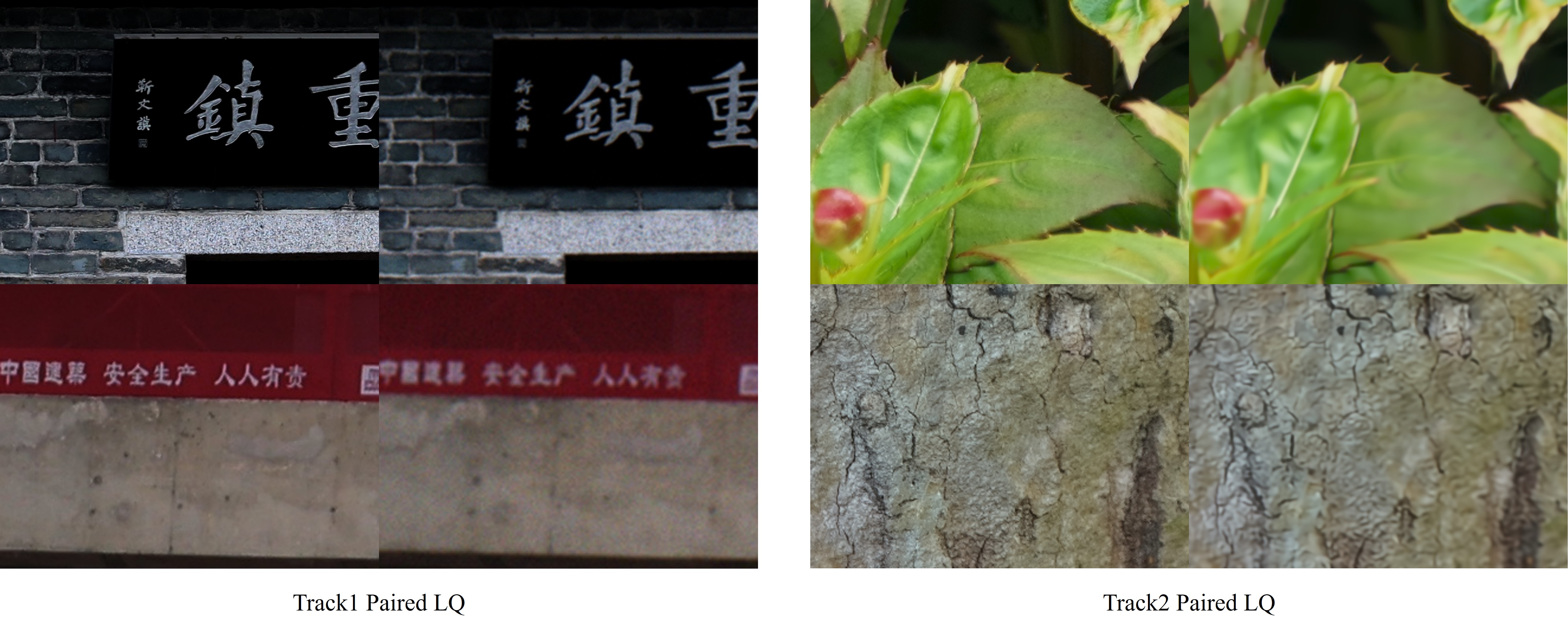} 
  \caption{Example Paired data for each track we have provided.} 
  \label{data_pair} 
  \vspace{2em}
\end{figure*}

\begin{figure*}
  \centering
  \includegraphics[width=1\textwidth]{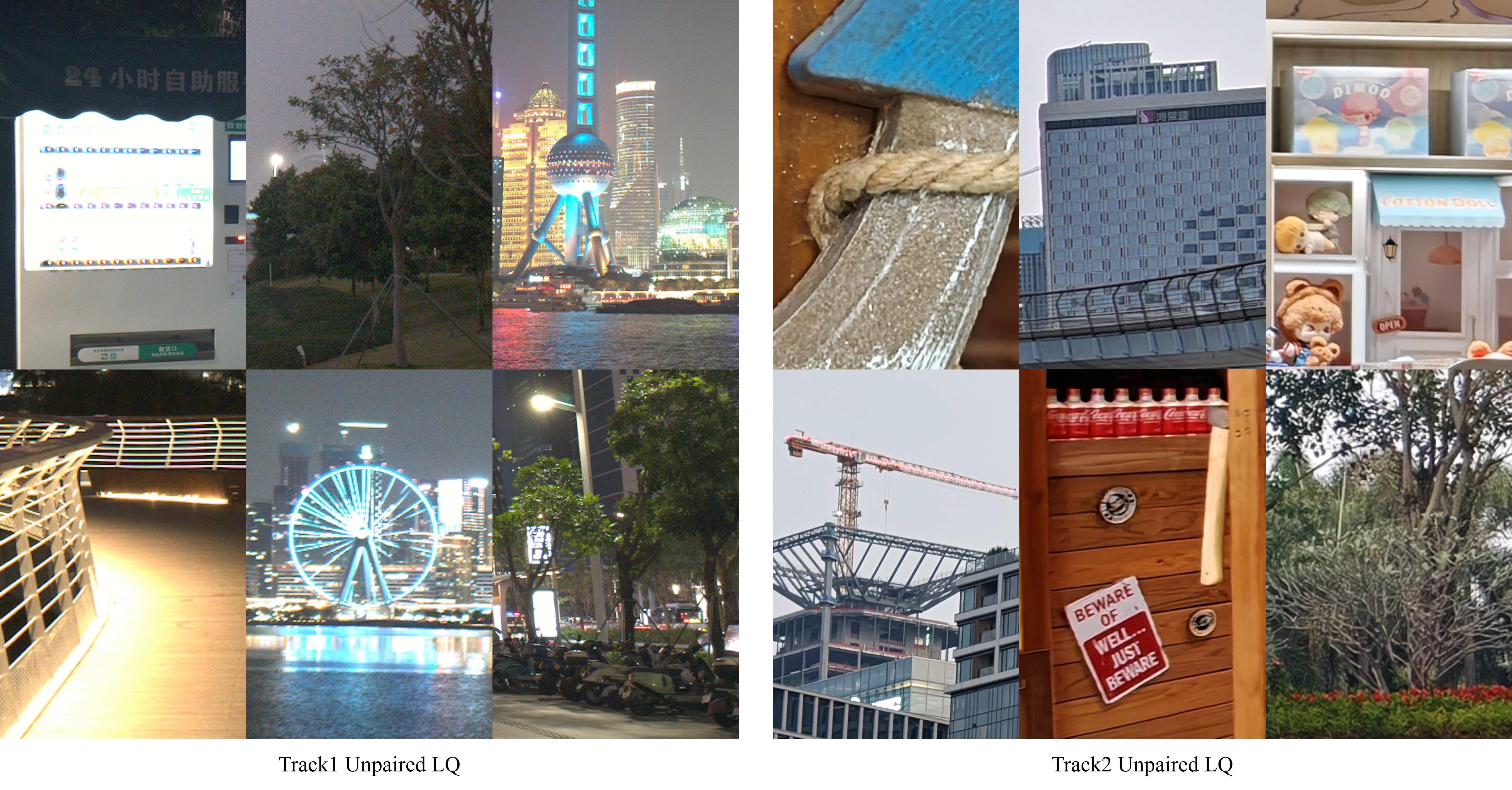} 
  \caption{Example Unpaired data for each track we have provided.} 
  \label{data1_unpair} 
  \vspace{2em}
\end{figure*}

To facilitate the model validation, we first provide some paired data in the following scenarios, where the R-GTs are of high quality and the input images are constructed via the real calibrated parameters and are processed via the real camera image signal processor (ISP).

In specific, in track one, \ie, the low-light joint denoising and demosaicing (JDD) task, we first calibrate the noise distribution of the camera sensor and the point spread function (PSF) of the camera lens, and then simulate the input raw images by applying the noise and the PSF in different intensities. In this way, we can get the paired data with both a high-quality R-GT (captured by a high-end DSLR) and a low-quality input that is as close to the real inputs as possible. For participants unacquainted with the raw-related tasks, we also provide the basic raw image processing codes for reference. We select sensors and lenses that are commonly applied in recent flagship smartphones (whose test data are also provided in phase 1 and phase 3), and the calibration of noise and the approximate PSF is also given for participants to simulate the training data. All data in track one can be available now at \url{https://drive.google.com/drive/folders/1Mgqve-yNcE26IIieI8lMIf-25VvZRs_J}.

In track two, \ie, the image detail enhancement/generation task, we first simulate the input raw images as in track one. Then, we process the raw inputs by applying a specific JDD model and a real camera ISP, which conducts tone mapping operations. The JDD model and the tone mapping operations that are key to this pipeline are from the real camera ISP operating in the real flagship smartphones. All data in track two can be available now at \url{https://drive.google.com/drive/folders/1UB7nnzLwqDZOwDmD9aT8J0KVg2ag4Qae}. Examples of two tracks can be found in Fig.~\ref{data_pair}.

We use these data to calculate the full-reference metrics to partially measure the effectiveness of the algorithms and screen the top performers in the early stage.

\subsubsection{Data without R-GT}
In many practical scenarios, the R-GT is very difficult to collect, and the image restoration performance is hard, if impossible, to be measured by full-reference metrics. In this challenge, we provide the data with the following commonly encountered issues in practice.

In track one, we provide raw inputs captured by real camera sensors in a single frame, without any postprocessing and keeping the original noise distribution. The test data covers various scenes (portrait/non-portrait, indoor/outdoor, and so on) and light conditions (different ISO and shutter). More attention should be paid to the following two problems in this track:

\noindent\textbf{Smoothed details and textures}. Limited by hardware and on-chip computing power, images captured by smartphone cameras often face a trade-off between noise/artifact reduction and details/texture richness, impacting the visual quality. The participants are encouraged to keep as rich high-frequency details as possible, to avoid over-smoothed results.

\noindent\textbf{Low-frequency color noise/blocks/bandings}. The low SNR of the input in mobile phone cameras demands a heavy denoising algorithm to output a clean image. However, due to factors such as computing power and storage, the bit-width of the ISP system is limited. When transitioning from the linear domain to the nonlinear domain, visual color noise, blocks, and bandings often appear.

In track two, we provide sRGB images that are captured by real flagship smartphones (the same as the one that provides the paired data). As in track one, the test data cover various scenes (portrait/non-portrait, indoor/outdoor, text, different distances, and so on) and light conditions (different ISO and shutter). More attention should be paid to the following two problems in this track:

\noindent\textbf{Realistic high-frequency detail enhancement or generation}. Limited by camera sensors and on-chip computing power, images captured by smartphone cameras always lack high-frequency details. The participants are encouraged to enhance or generate realistic details to further improve the quality of the image.

\noindent\textbf{Texture adhesion in super-resolution}. In the telephoto mode (e.g., equivalent focal length larger than 230mm), shooting small structures from a distance is an important yet highly challenging task. Texture adhesion greatly deteriorate the user experience.

Examples of two tracks can be found in Fig.~\ref{data1_unpair}.


\subsection{Evaluation Measures}

We evaluate the effectiveness of the models with both quantitative measures and subjective evaluation. We also consider the efficiency of the models, including the number of parameters, the FLOPs, and the running time. 

\subsubsection{Quantitative Measure}
\label{quantitativemeasure}

Following prior arts and the 1st RAIM, we employ the PSNR, SSIM, LPIPS, DISTS, and NIQE measures to evaluate the models quantitatively by using the data with R-GT. The evaluation score is computed as follows\footnote{The script of this measure is available at\url{https://sbox.myoas.com/outpublish.html?code=A51c91a521084c0b0\#view}}:

\noindent $SCORE = 20\times \frac{PSNR}{50}+15\times \frac{SSIM-0.5}{0.5}+20\times (1-\frac{LPIPS}{0.4})+40\times (1-\frac{DISTS}{0.3})+30\times (1-\frac{NIQE}{10})$.

\subsubsection{Subjective Evaluation}
\label{qualitativemeasure}

For the test data without R-GT, we judge the perceptual quality of the restored results by visual inspection. Specifically, we invite 10 experienced practitioners and conduct comprehensive user studies for each track. The following features should be considered in the evaluation:

\noindent\textbf{Textures and details}. The restored image should have fine and natural textures and details.

\noindent\textbf{Noise}. Noise, especially color noise, should be eliminated. Some luminance noise should be kept to avoid over-smoothness in flat areas.

\noindent\textbf{Artifacts}. Various artifacts, such as worm-like artifacts, color blocks, bandings, over-sharpening, and so on, should be reduced as much as possible.

\noindent\textbf{Fidelity}. The restored image should be loyal to the given input.

 More details have been discussed during the competition with all participants by referencing specific images and model outputs.

\subsubsection{Efficiency}
\label{efficiency}
In Phase 2 and Phase 3, we consider the efficiency of the model as an important evaluation metric. 
To ensure objectivity and comparability in efficiency evaluation, we establish a standardized efficiency scoring system based on three key factors: parameter count, computational complexity, and inference time.

\noindent\textbf{Parameter Count}. The total number of parameters in the model, measured in millions (M).

\noindent\textbf{Computational Complexity (FLOPs)}. The number of floating-point operations required for a single forward pass on a $1024 \times 1024$ input image, measured in billions (B). 

\noindent\textbf{Inference Time}. The average forward-pass time per image (in milliseconds) when running the full test set on the same hardware.

For each of the three efficiency metrics—parameter count, FLOPs, and inference time—models are ranked in ascending order, where a lower value indicates better performance. A maximum of 10 points is assigned to the top-ranked model, and the remaining models receive proportionally worse scores based on their rank. The score interval between ranks is calculated as $10/(N-1)$, where 
$N$ is the total number of submitted models. Models with identical metric values share the same score, and subsequent ranks are skipped accordingly to maintain fairness. The final efficiency score is the sum of the scores from three metrics.

\subsection{Phases}

\subsubsection{Phase 1: Model Design and Tuning}
In this phase, participants can analyze the given data and tune their models accordingly. We provided:

\begin{itemize}
\item 50 pairs of paired data in high resolution (i.e., input with reference ground-truth (R-GT)), which can be used to tune the models based on the quantitative measures;

\item 50 images without R-GT, which can be used to tune the model according to visual perception.
\end{itemize}

\subsubsection{Phase 2: Online Feedback}
In this phase, participants can upload their results and get official feedback. We provide:

\begin{itemize}
\item The input low-quality images of another 100 pairs of paired data. 
\end{itemize}

\noindent Only the low-quality input images are provided, and the participants can upload the restoration results to the server and get the quantitative scores online. Users can also upload their results of the images without R-GT provided in Phase 1 to seek feedback. The organizers will provide feedback to a couple of teams that get the highest quantitative scores for the images with R-GT.

\subsubsection{Phase 3: Final Evaluation}
In this phase, we provide the following files for each track:

\begin{itemize}
\item Another 50 images without R-GT for subjective evaluation.
\end{itemize}

\noindent We select the top ten teams according to the quantitative score of the 100 images with R-GT in Phase 2 for each track, and then arrange a comprehensive user study on their results on the above 50 images without R-GT. The final ranks of the ten teams will be decided based on the quantitative scores in phase 2, the efficiency of the model (including \#Parameters, FLOPs, and Running Time of the final model), and the subjective user study results (the weight will be given before the beginning of phase 3).

\subsection{Awards}

The following awards of each track are provided \textbf{for each track}:

\begin{itemize}
\item One first-class award (i.e., the champion) with a cash prize of \textbf{US\$1000};

\item Two second-class awards with cash prizes of \textbf{US\$500 each};

\item Three third-class awards with cash prizes of \textbf{US\$200 each}.
\end{itemize}

\subsection{Important Dates}

\begin{itemize}
\item 2025.02.07: Released data of phase 1. Phase 1 began;
\item 2025.02.24: Released data of phase 2. Phase 2 began;
\item 2025.03.14: Released data of phase 3. Phase 3 began;
\item 2025.03.28: Phase 3 results submission deadline;
\item 2025.04.03: Final rank announced.
\end{itemize}
\section{Challenge Results}

In total, Tracks 1 and 2 of the challenge received 156 and 141 registrations, respectively. During Phase 2, 22 participants submitted 324 results in Track 1, while 29 participants submitted 362 results in Track 2. In Phase 3, we invited the top 7 and 12 teams from Phase 2 of Tracks 1 and 2, respectively. A brief overview of the methods used by the top 6 teams is presented in Section~\ref{teamsandmethods}, and detailed information about all participating teams can be found in Section~\ref{appendix}.

\subsection{Phase 2: quantitative comparison on paired data with R-GT}
In phase 2, we got submissions from 22 and 29 teams in tracks 1 and 2, where the quantitative results of top-ranked teams are shown in Table~\ref{table_final_result} and Table~\ref{table_final_result_2}. The evaluation measure is described in Section \ref{quantitativemeasure}.

\begin{table}\scriptsize
\renewcommand{\arraystretch}{1.2}
\caption{Result of phases 2 and 3 in track 1, as well as final scores and ranks, where the `EScore' and `UScore' denote the efficiency score and user study score, respectively, and the `P2' and `P3' denote the phase 2 and 3, respectively. We only show teams that participated in phase 3.}
\label{table_final_result}
\centering
\resizebox{\linewidth}{!}{
\begin{tabular}{lccccc}
\toprule
Team& Score in P2&EScore in P3&UScore in P3&Final Score&Rank\\
\midrule
MiAlgo&100.89&15&106&112.18&1\\
IID-AI&106.16&8.34&84&91.30&2\\
PolyU-AISP&89.32&16.67&32&55.34&3\\
TongJi-IPOE&121.16&30&0&48.23&4\\
NJUST-KMG&111.79&25&6&46.89&5\\
Xianggkl&88.38&0&25&36.54&6\\
WANGTY&107.27&10&1&30.21&7\\
\bottomrule
\end{tabular}}
\end{table}

\begin{table}\scriptsize
\renewcommand{\arraystretch}{1.2}
\caption{Result of phases 2 and 3 in track 2, as well as final scores and ranks, where the `EScore' and `UScore' denote the efficiency score and user study score, respectively, and the `P2' and `P3' denote the phase 2 and 3, respectively. We only show teams that participated in phase 3.}
\label{table_final_result_2}
\centering
\resizebox{\linewidth}{!}{
\begin{tabular}{lccccc}
\toprule
Team& Score in P2&EScore in P3&UScore in P3&Final Score&Rank\\
\midrule
NulltoZero&106.44&20.9&27.45&113.26&1\\
TeleAI-Vision&97.33  & 5.45  & 29.18 & 111.83& 2\\
TongJi-IPOE&116.75 & 4.54  & 24.64 & 110.54 & 3\\
NJUST-KMG&118.51 & 4.54  & 23.36 & 107.38 & 4 \\
iAM\_IR&116.55 & 5.45  & 22.55 & 105.49  & 5\\
MiAlgo&99.01  & 8.18  & 21.91 & 102.41 & 6 \\
WANGTY&100.04 & 5.45  & 26.55 & 101.53  & 7\\
Alchemist&91.17  & 0     & 29    & 101.38   & 8\\
Hitsz-Go&89.31  & 2.73  & 26.64 & 100.36& 9\\
ICANT-IR&95.26  & 0.91  & 26.82 & 95.26  & 10\\
TACO\_SR&83.37  & 3.64  & 22.91 & 81.67 & 11 \\
webbzhou&108.7  & 7.28  & 18.45 & 78.15   & 12\\
\bottomrule
\end{tabular}}
\end{table}

\begin{figure*}
  \centering
  \includegraphics[width=0.95\textwidth]{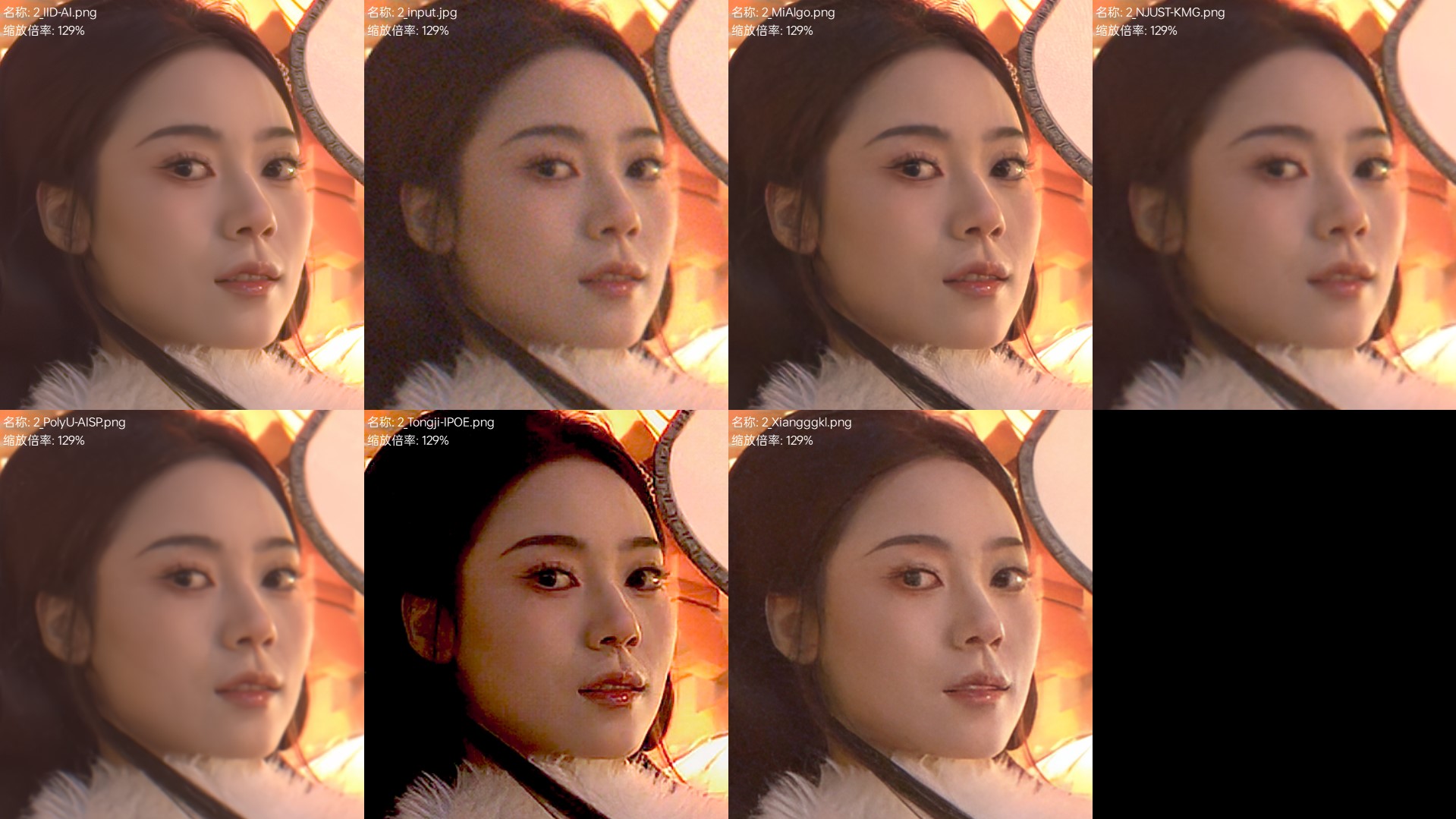} 
  \caption{Visual comparisons between the input LQ image and the results from the top 6 participating teams in Track 1.} 
  \label{fig:track_1} 
  \vspace{2em}
\end{figure*}

\begin{figure*}
  \centering
  \includegraphics[width=0.95\textwidth]{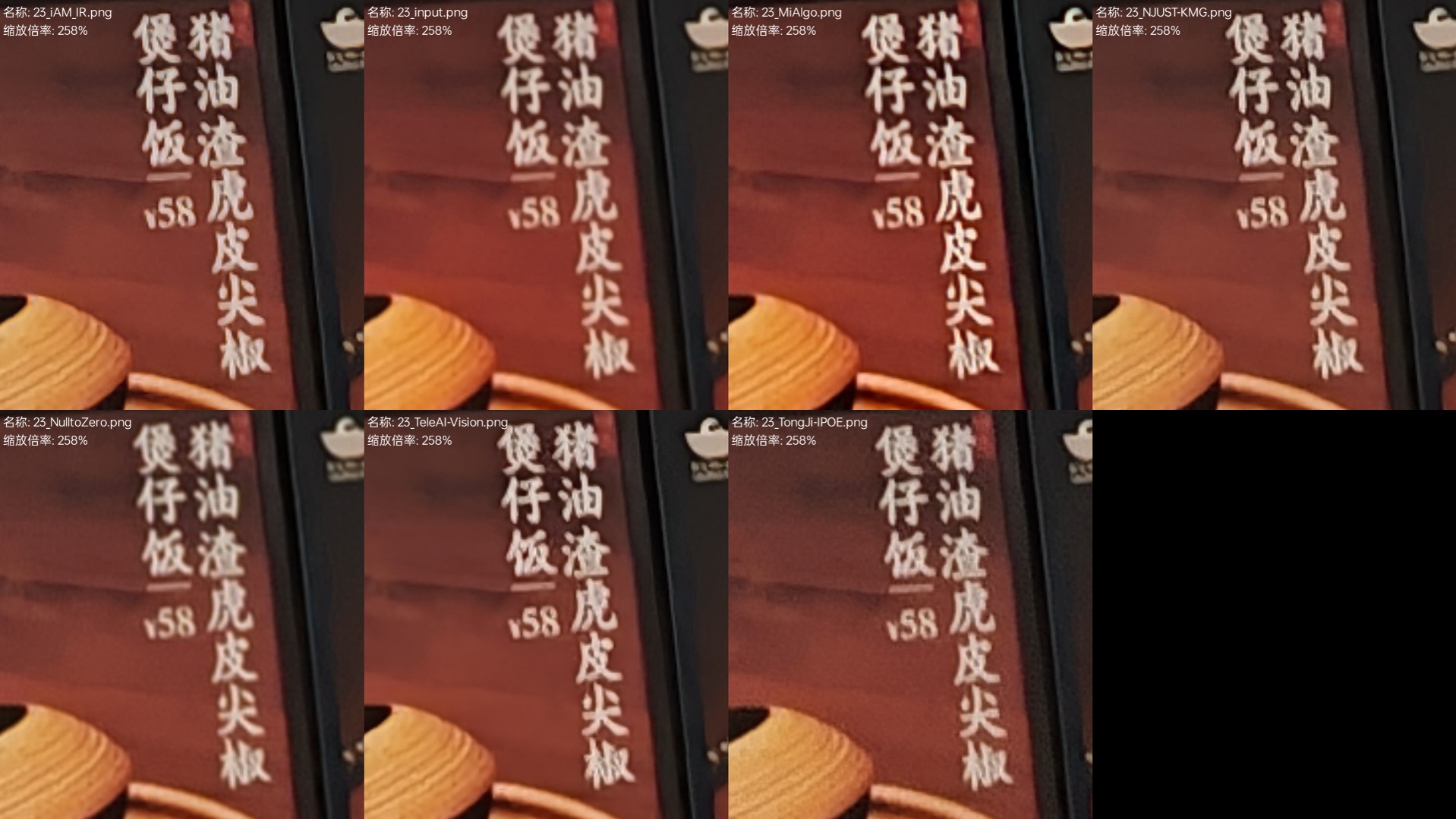} 
  \caption{Visual comparisons between the input LQ image and the results from the top 6 participating teams in Track 2.} 
  \label{fig:track_2} 
\end{figure*}

\subsection{Phase 3: qualitative comparison on unpaired data}

In stage 3, the efficiency score is calculated as Section \ref{efficiency}. In addition, we invite 20 low-level vision-related students/engineers, who are required to select the top three results of each of the 50 samples. They follow a unified principle as demonstrated in Section~\ref{qualitativemeasure} and the feedback to individual participants. The team information is hidden and the results are randomly shuffled to make fair comparisons. By checking the results of each scorer, we found their opinions are similar so the results are valid. The final score $S_{final}$ is calculated by 
\begin{equation}
S_{final} = 0.4 \times S_2 + 0.6 \times S_3^n,
\end{equation}
where $S_2$ indicates the score in phase 2 and $S_3^n$ denotes the normalized score in phase 3. 

For calculating $S_3^n$, we first calculate the score in phase 3, \ie, $S_3$, where the team is rewarded with 3 points when selected to be top 1, 2 points for the top 2, and 1 point for the top 3. The scores are averaged by 18. Then, we calculate $S_3^n$ by
\begin{equation}
{S_3^n}_{team i} = \frac{S_{3_{team i}} - min(S_3)}{max(S_3) - min(S_3)}.
\end{equation}

We then show some example visual comparisons in Fig.~\ref{fig:track_1}, and Fig.~\ref{fig:track_2}. All visual results in phase 3 are available at \url{https://drive.google.com/drive/folders/1-N_5Ebd0iA7KOzOWqajM7ZvNgcl1KkKH?usp=share_link}.






\section{Teams and Methods}
\label{teamsandmethods}
Due to space limitations, we describe the participating teams and their proposed methods for the two tracks in the supplementary material.

\section{Acknowledgments}

This work was partially supported by the Humboldt Foundation. We thank the NTIRE 2025 sponsors: ByteDance, Meituan, Kuaishou, and University of Wurzburg (Computer Vision Lab).

\section{Appendix: Teams and affiliations}
\label{appendix}

\textbf{NTIRE 2025 Team}
\flushleft

\noindent\textit{\textbf{Challenge:}} 

\noindent NTIRE 2025 Restore Any Image Model (RAIM) in the Wild

\noindent\textit{\textbf{Organizers:}}

\noindent Jie Liang$^1$ (liang27jie@gmail.com)

\noindent Qiaosi Yi$^{1,2}$ (qiaosi.yi@connect.polyu.hk)

\noindent Zhengqiang Zhang$^{1,2}$ (22040257r@connect.polyu.hk)

\noindent Shuaizheng Liu$^{1,2}$ (shuaizhengliu21@gmail.com)

\noindent Lingchen Sun$^{1,2}$ (22042099r@connect.polyu.hk)

\noindent Rongyuan Wu$^{1,2}$ (rong-yuan.wu@connect.polyu.hk)

\noindent Xindong Zhang$^{1}$ (17901410r@connect.polyu.hk)

\noindent Hui Zeng$^{1}$ (cshzeng@gmail.com)

\noindent Prof. Lei Zhang$^{1,2}$ (cslzhang@comp.polyu.edu.hk)

\noindent Prof. Radu Timofte$^3$ (radu.timofte@uni-wuerzburg.de)

\noindent\textit{\textbf{Affiliations:}}

\noindent $^1$ OPPO Research Institute

\noindent $^2$ The Hong Kong Polytechnic University 

\noindent $^3$ Computer Vision Lab, University of W\"urzburg, Germany

~\\

\textbf{Track 1}

\noindent\textit{\textbf{Team name:}} MiAlgo

\noindent\textit{\textbf{Members:}}  
Tianyu Hao (haotianyu@xiaomi.com), Yuhong He, Ruoqi Li, Yueqi Yang, Yibin Huang, Meng Zhang, Jingyuan Xiao, Chaoyu Feng, Xiaotao Wang, Lei Lei

\noindent\textit{\textbf{Affiliations:}} Xiaomi Inc., China

~\\

\noindent\textit{\textbf{Team name:}} IID-AI

\noindent\textit{\textbf{Members:}}  
Lin Wang (wanglin@intellindust.com), Xi Shen, Cheng Li, Xiuxin Li, Peizhe Ru

\noindent\textit{\textbf{Affiliations:}} Intellindust Information Technology Co., Ltd, China

~\\
\noindent\textit{\textbf{Team name:}} PolyU-AISP

\noindent\textit{\textbf{Members:}}  
Zhe Xiao (xiao-zhe.xiao@connect.polyu.hk), Yushen Zuo, Zongqi He, Kin-Chung Chan, Hanmin Li, Hao Xie, Zihang Lyu, Cong Zhang, Jun Xiao, Kin-Man Lam

\noindent\textit{\textbf{Affiliations:}}  
The Hong Kong Polytechnic University, Sun Yat-sen University

~\\

\noindent\textit{\textbf{Team name:}} TongJi-IPOE

\noindent\textit{\textbf{Members:}}  
Pengzhou Ji (jipengzhou@tongji.edu.cn), Xiong Dun, Zhiyuan Ma, Xuanyu Qian, Jian Zhang, Xuquan Wang, Zhanshan Wang, Xinbin Cheng

\noindent\textit{\textbf{Affiliations:}}  
Institute of Precision Optical Engineering, School of Physics Science and Engineering, Tongji University, China

~\\
\noindent\textit{\textbf{Team name:}} NJUST-KMG

\noindent\textit{\textbf{Members:}}  
Shu-Peng Zhong (zspnjlgdx@gmail.com), Kai Feng, Xiao-Long Yin, Yang Yang, Jie Liu

\noindent\textit{\textbf{Affiliations:}}  
Nanjing University of Science and Technology, China

~\\
\noindent\textit{\textbf{Team name:}} Xianggkl

\noindent\textit{\textbf{Members:}}  
Xiangming Wang (xmwang28@gmail.com), Jinchao Li, Haijin Zeng, Shiyang Zhou, Kai Feng, Yongyong Chen, Jingyong Su, Jie Liu

\noindent\textit{\textbf{Affiliations:}}  
Harbin Institute of Technology, Shenzhen (HITSZ), China

~\\

\noindent\textit{\textbf{Team name:}} WANGTY

\noindent\textit{\textbf{Members:}}  
Wang ZhanSheng (wangty537@gmail.com), Wu Chen, Zhang PengBo, Huang Jiazi

~\\

\textbf{Track 2}

\noindent\textit{\textbf{Team name:}}
NulltoZero

\noindent\textit{\textbf{Members:}}

Shu-Peng Zhong (zspnjlgdx@gmail.com)

Xiang Yang, Jing-Yuan Wang, Wei-Li Guo, Qing-Yuan Jiang

\noindent\textit{\textbf{Affiliations:}}

Nanjing University of Science and Technology

~\\

\noindent\textit{\textbf{Team name:}}
TeleAI-Vision

\noindent\textit{\textbf{Members:}}

Jiaqi Yan (jiaqi\_yan@smail.nju.edu.cn)

Shuning Xu, Xiangyu Chen

\noindent\textit{\textbf{Affiliations:}}

Institute of Artificial Intelligence (TeleAI), China Telecom; State Key Laboratory for Novel Software Technology, Nanjing University, Nanjing 210023, China

~\\

\noindent\textit{\textbf{Team name:}}
TongJi-IPOE 

\noindent\textit{\textbf{Members:}}

Pengzhou Ji (jipengzhou@tongji.edu.cn)

Xiong Dun, Ziyu Zhao, Wenhan Huang, Yujie Xing, Xuquan Wang, Zhanshan Wang, Xinbin Cheng

\noindent\textit{\textbf{Affiliations:}}

Institute of Precision Optical Engineering, School of Physics Science and Engineering, Tongji University

~\\

\noindent\textit{\textbf{Team name:}}
NJUST-KMG

\noindent\textit{\textbf{Members:}}

Sishun Pan (pansishun@njust.edu.cn)

Xiaolong Yin, Yang Yang

\noindent\textit{\textbf{Affiliations:}}

Nanjing University of Science and Technology

~\\

\noindent\textit{\textbf{Team name:}}
iAM\_IR

\noindent\textit{\textbf{Members:}}

Ce Wang (cewang@whu.edu.cn)

Zhenyu Hu, Wanjie Sun, Zhenzhong Chen

\noindent\textit{\textbf{Affiliations:}}

School of Remote Sensing and Information Engineering, Wuhan University

~\\

\noindent\textit{\textbf{Team name:}}
MiAlgo

\noindent\textit{\textbf{Members:}}

Yibin Huang (huangyibin@xiaomi.com)

Long Cai, Tianyu Hao, Lize Zhang, Shuai Liu, Yuhong He,  Guangqi Shao, Xiaotao Wang, Lei Lei

\noindent\textit{\textbf{Affiliations:}}

Xiaomi Inc., China

~\\

\noindent\textit{\textbf{Team name:}}
WANGTY

\noindent\textit{\textbf{Members:}}

ZhanSheng Wang (wangty537@gmail.com)

Chen Wu, PengBo Zhang, Jiazi Huang

~\\

\noindent\textit{\textbf{Team name:}}
Alchemist

\noindent\textit{\textbf{Members:}}

Haobo Liang (11230752@stu.lzjtu.edu.cn)

Jiajie Jing, Junyu Li, Quanjun Zhang, Feiyu Tong

\noindent\textit{\textbf{Affiliations:}}

1. School of Electronics and Information Engineering, Lanzhou Jiaotong University

2. School of Mathematics and Physics, Lanzhou Jiaotong University

~\\

\noindent\textit{\textbf{Team name:}}
Hitsz-Go

\noindent\textit{\textbf{Members:}}

Zhenghao Pan (pzh20000803@gmail.com)

Benteng Sun, Haijin Zeng, Yongyong Chen

\noindent\textit{\textbf{Affiliations:}}

Harbin Institute of Technology (Shenzhen)

~\\

\noindent\textit{\textbf{Team name:}}
ICANT-IR

\noindent\textit{\textbf{Members:}}

Jinjian Wu (wujinjian@mail.nwpu.edu.cn)

Jiaqi Tang, Xinjie Zhang, Bowen Fu, Jianmin Chen, Yuduo Bian, Lei Zhang, Wei Wei

\noindent\textit{\textbf{Affiliations:}}

Northwestern Polytechnical University

~\\

\noindent\textit{\textbf{Team name:}}
TACO\_SR

\noindent\textit{\textbf{Members:}}

Yushen Zuo$^1$ (zuoyushen12@gmail.com)

Mingyang Wu$^2$, Renjie Li$^2$, Shengyun Zhong$^3$, Zhengzhong Tu$^2$

\noindent\textit{\textbf{Affiliations:}}

1. The Hong Kong Polytechnic University

2. Texas A\&M University

3. Northeastern University

~\\

\noindent\textit{\textbf{Team name:}}
webbzhou

\noindent\textit{\textbf{Members:}}

Yuanbo Zhou (webbozhou@gmail.com)

Wei Deng,  Qinquan Gao, Tong Tong

\noindent\textit{\textbf{Affiliations:}}

Fuzhou University,
Imperial Vision Technology
\section{Teams and Methods}
\label{teamsandmethods}

This section briefly describes the participating teams and their proposed methods for the two tracks. We only provide the methods of the top six teams of each track.

\subsection{Track 1}
\subsubsection{Team MiAlgo}
Team MiAlgo proposed a two-stage raw image restoration pipeline. Their method integrates transformer-based and GAN-based models for joint denoising, demosaicking, and detail enhancement, with robustness to noise, exposure, and sensor defects.

\begin{figure}[t]
	\centering
	\includegraphics[width=0.5\textwidth]{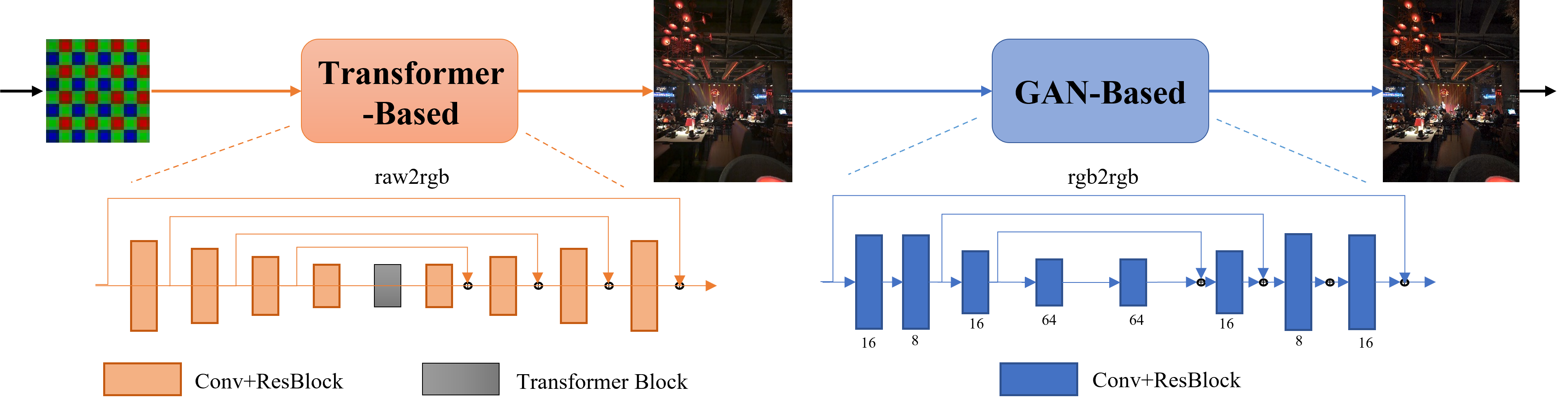}
	\caption{The two-stage pipeline proposed by team MiAlgo.}
	\label{fig:0001_method}
\end{figure}

\noindent\textbf{Architecture.} As shown in Fig.~\ref{fig:0001_method}, the proposed method consists of a two-stage end-to-end pipeline. Stage~1 uses Restormer to jointly perform denoising and demosaicking. Its transformer-based self-attention mechanism enables modeling of long-range dependencies and better preservation of semantic features. Stage~2 applies a GAN model for texture enhancement, where the discriminator follows the Real-ESRGAN~\cite{wang2021real} design. The generator in this stage is initialized with Stage~1 parameters to boost convergence and performance. In Phase~3, the team compressed and distilled their Phase~2 model using a lightweight UNet architecture enhanced with basic transformer blocks and two Restormer modules in the bottleneck. MWRCAN~\cite{ignatov2020aim} is used as the Stage~2 generator to enable efficient multi-scale restoration.

\noindent\textbf{Data Augmentation.} The team used both the official paired dataset and an internal ultra-high-resolution dataset (4K–6K) consisting of 1200 images, including 1000 general scenes and 200 night portraits. The data degradation pipeline includes gamma correction, AWB gain removal, CCM adjustment, blurring, noise, downsampling, and ISO-based noise augmentation using darkening and dgain transformation. To simulate sensor defects, random Bayer pattern defect augmentation was also applied.

\noindent\textbf{Training Details.} The training was conducted in two stages. Stage~1 was trained with $\mathcal{L}_1$ loss for 300K iterations at a resolution of $256 \times 256$, followed by 100K iterations at $512 \times 512$ using a loss function composed of L2 + 0.1 Perceptual + 0.1 LPIPS + 0.1 SSIM. Stage~2 was trained with a composite loss of L2 + 0.1 Perceptual + 0.01 GAN + 4 LPIPS, using a learning rate of $1 \times 10^{-5}$. The models were implemented in PyTorch and trained on 8 NVIDIA A100 GPUs.

\noindent\textbf{Performance and Efficiency.} The method achieved the highest score in Phase~2 with a PSNR-based score of 92.56. The Phase~2 model has 52.24M parameters and 5351.47 GFLOPs (for $1024 \times 1024$ input), with an average inference time of 2651~ms per image. The Phase~3 model is significantly lighter, with 4.62M parameters and 372.09 GFLOPs, and runs at 320~ms per image (3072×4096 resolution) on A100.

\noindent\textbf{Robustness and Generality.} The proposed solution demonstrates strong robustness in real-world scenarios, especially in the presence of extreme exposure and Bayer pattern defects. The lightweight design of Phase~3 ensures practical deployment with minimal compromise in performance.

\noindent\textbf{Additional Information.} No external or pre-trained models were used. No ensemble or fusion strategies were applied. The method is novel and has not been published. The team does not plan to submit a paper for NTIRE 2025. They suggest releasing the evaluation metrics earlier in future competitions for better tuning and transparency.

\subsubsection{Team IID-AI}
Team IID-AI employed a data-centric strategy and adopted XRestormer~\cite{chen2024comparative} as the backbone network to address the domain gap between training and testing data. The model is trained with a carefully designed multi-loss function to achieve high-quality raw image enhancement.

\noindent\textbf{Data Synthesis.} Initial experiments trained on Phase~1 paired data showed strong performance on Phase~2 paired testing, but significant degradation on Phase~3 unpaired data. The root cause was identified as a domain shift: DSLR-captured (16-bit, daytime) training data versus smartphone-captured (10-bit, nighttime) testing data. To mitigate this gap, over 700 clean RAW images were captured using smartphones under low-ISO and short-exposure conditions to simulate noise-free ground truth. A degradation pipeline was then applied, including demosaicing, Gaussian blurring (based on PSF parameters), downsampling, and synthetic noise generation using ISO 1600, 3200, and 6400-level noise. GT sRGB images were generated using AAHD demosaicing for improved edge sharpness and texture fidelity. This pipeline effectively supports joint denoising and demosaicking.

\noindent\textbf{Network Design.} XRestormer~\cite{chen2024comparative} was selected as the backbone for its proven effectiveness in low-level vision tasks and efficient computation. To adapt it to the raw-to-RGB task, the final convolutional layer was modified to output 12 channels, followed by a pixelshuffle layer for resolution doubling and RGB conversion. The input–output skip connection was removed due to channel mismatch.

\noindent\textbf{Loss Functions.} A combination of L1 reconstruction loss~\cite{lai2018fast}, perceptual loss~\cite{zhang2018unreasonable, johnson2016perceptual}, and FFT-based frequency loss~\cite{krawczyk2023artifact} was used. While reconstruction and perceptual losses ensured good quantitative and perceptual quality, checkerboard artifacts were observed in dark regions of Phase~3 test images. This was addressed by adding a frequency-domain loss term. The overall training loss consists of an L1 loss, an FFT loss weighted by 300, and an LPIPS loss weighted by 0.3.

\noindent\textbf{Training Details.} The model was implemented using PyTorch and trained on an NVIDIA RTX 4090 GPU. The optimizer was Adam, with a batch size of 2 and a patch size of 512. Training was conducted for 70K iterations with a learning rate of $1 \times 10^{-4}$, followed by 30K iterations at $1 \times 10^{-5}$.

\subsubsection{Team PolyU-AISP}

Team PolyU-AISP proposed a lightweight restoration network based on the NAFNet~\cite{chen2022simple} architecture. The method adopts a UNet-like design with layer normalization, convolution, and channel attention in each block. To handle high-resolution inputs, the model processes images in overlapping tiles and fuses them at the output, achieving both computational efficiency and restoration quality.

\begin{figure}[htbp]
\centering
\includegraphics[scale = 0.26]{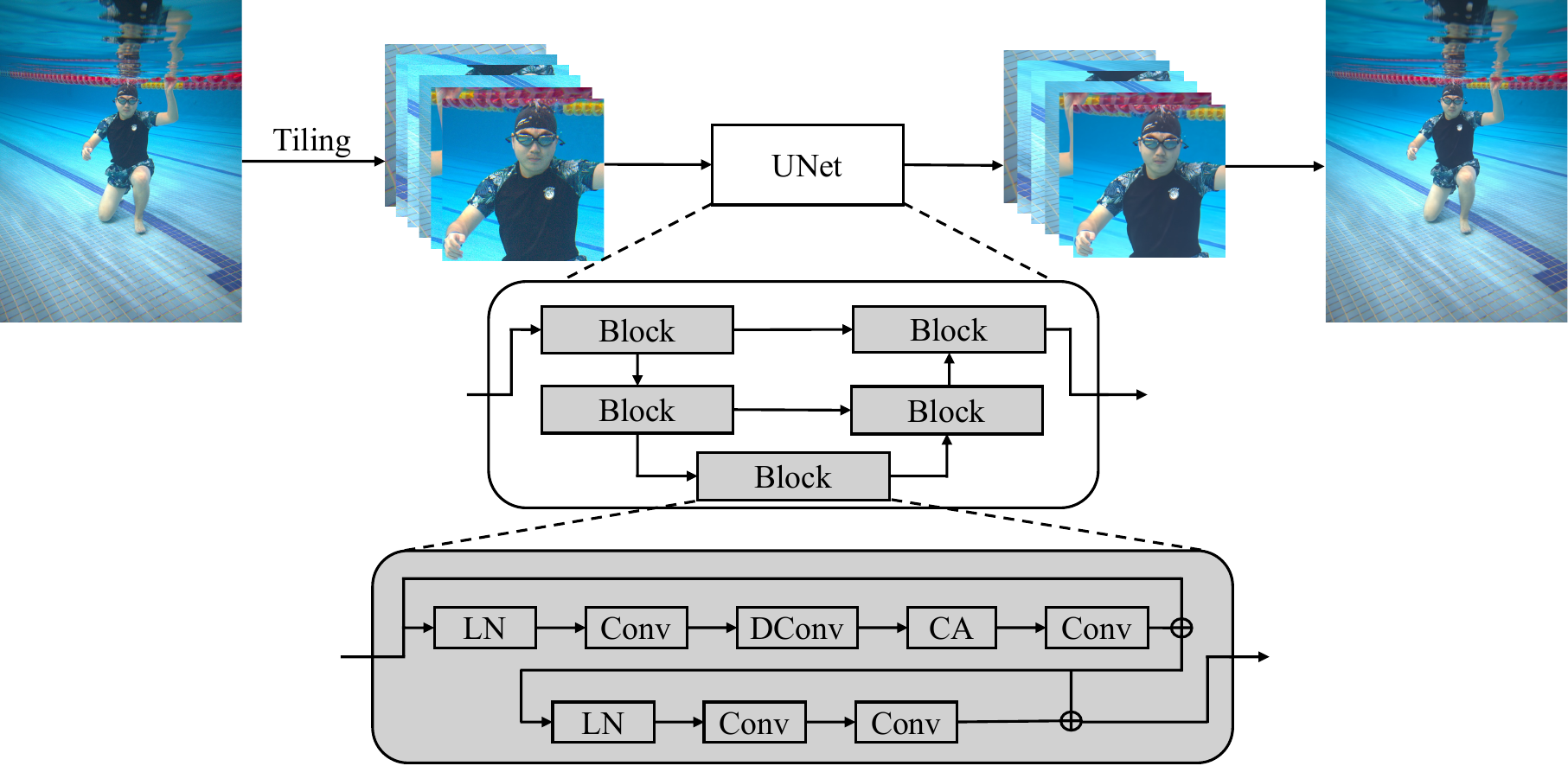}
\caption{
Pipeline of Team POLYU-AISP. LN: Layer Normalization, Conv: 1×1 Convolution, DConv: Depthwise Convolution, CA: Channel Attention (as in NAFNet~\cite{chen2022simple}). Certain details, such as activation functions and element-wise operations, are omitted for clarity.
}
\label{fig:pipeline_POLYU-AISP}
\end{figure}

\noindent\textbf{Architecture.}  
As illustrated in Fig.~\ref{fig:pipeline_POLYU-AISP}, the proposed pipeline is a compact UNet-style network built upon NAFNet. Each block integrates layer normalization, 1×1 convolution, depthwise convolution, and channel attention. The input image is divided into $2048 \times 2048$ overlapping tiles with 64-pixel padding. These tiles are processed independently and then fused using weighted averaging to ensure smooth transitions across tile boundaries. The final model contains 29M parameters, offering a good trade-off between performance and model size.

\noindent\textbf{Training Details.}  
The model is trained on the SIDD dataset~\cite{abdelhamed2018high}, following the protocol in HiNet~\cite{chen2021hinet}. The training patch size is $256 \times 256$, and the network is optimized with the Adam optimizer~\cite{kingma2014adam}. The learning rate starts from $10^{-3}$ and decays to $10^{-7}$ using cosine annealing over 200K iterations. A tile width of 32 is used during training.

\noindent\textbf{Testing and Inference.}  
During inference, the input is tiled into $2048 \times 2048$ overlapping patches with 64 pixels of overlap. Each tile is processed through the network, and the outputs are fused via weighted averaging. The average runtime per image is approximately 2.8 seconds on an NVIDIA RTX 4090 GPU. The method requires only 3K GFLOPs to process a full $4096 \times 3072$ image, demonstrating high computational efficiency.

\subsubsection{Team TongJi-IPOE}

Team TongJi-IPOE proposed a lightweight and efficient joint raw image denoising and demosaicing solution, named B2FNet (Branching to Fusion Network), designed for low-light image processing. Inspired by the low-light pipeline from Chen et al.~\cite{chen2018learning}, the method explicitly separates and processes the green and red/blue channels in the Bayer pattern, followed by a fusion stage to generate the final sRGB image.

\begin{figure}[t]
    \centering
    \includegraphics[width=80mm]{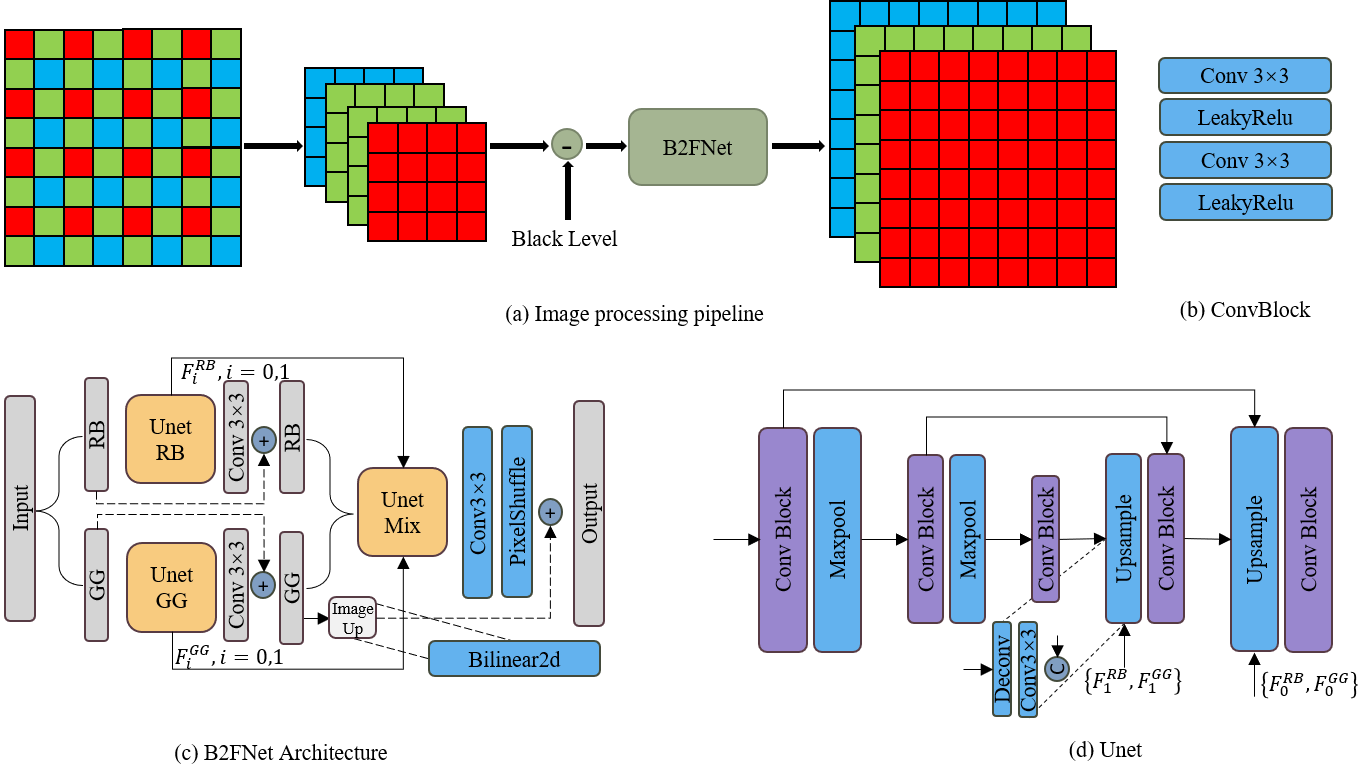}
    \caption{Overview of the Team TongJi-IPOE.}
    \label{fig:TongJi-IPOE_diagram}
\end{figure}

\noindent\textbf{Architecture.}  
As shown in Fig.~\ref{fig:TongJi-IPOE_diagram}, B2FNet is a three-stage UNet-based structure. The input raw image is first converted to a four-channel RGGB image via demosaicing. Two lightweight UNets are deployed to restore the GG and RB channels separately. Their outputs are then fused by another UNet to generate the final sRGB output. Each UNet consists only of $3 \times 3$ convolutional layers, max-pooling for downsampling, and deconvolution for upsampling. To avoid gradient vanishing, feature maps from the encoder stages of the two-branch UNets are introduced into the decoder of the fusion UNet.

\noindent\textbf{Processing Pipeline.}  
Given a raw input of size $\mathbb{R}^{H \times W \times 1}$, it is first demosaiced into a four-channel RGGB image $\mathbb{R}^{\frac{H}{2} \times \frac{W}{2} \times 4}$. This is followed by dark-level correction and processed by B2FNet to generate a final sRGB image of size $\mathbb{R}^{H \times W \times 3}$.

\noindent\textbf{Training Details.}  
The model is trained solely on the provided dataset, using PyTorch on a single NVIDIA RTX 3090Ti GPU. The optimizer is AdamW with $\beta_1 = 0.9$, $\beta_2 = 0.999$. The training is conducted in three stages: 92K iterations with a fixed learning rate of $1 \times 10^{-4}$, followed by 208K iterations using cosine annealing to decay the learning rate to $1 \times 10^{-6}$, and finally 300K iterations of fine-tuning at $1 \times 10^{-5}$. The training patch size is $128 \times 128$.

\noindent\textbf{Testing and Efficiency.}  
At inference, raw inputs are first demosaiced to RGGB format, corrected for dark levels, and passed through B2FNet for sRGB conversion. The model contains only 0.39M parameters and requires 23.45 GFLOPs for an input size of $1024 \times 1024$. Average inference time is 8.3 ms on an NVIDIA A100 GPU.

\noindent\textbf{Training Details.}  
The method is implemented in PyTorch and runs on a single RTX 3090Ti GPU. The pipeline is simple and compact, and suitable for real-time deployment. The full training and fine-tuning process took approximately 24 hours. Given its low complexity, the method holds potential for edge deployment with further optimization.

\subsubsection{Team NJUST-KMG}

Team NJUST-KMG proposed a novel restoration framework for low-light raw image denoising and demosaicing, named DWT-Enhanced Hybrid Networks. The approach combines a hybrid attention mechanism with multi-scale feature processing via discrete wavelet transforms (DWT), and is trained via a two-stage pipeline to jointly optimize noise removal and detail preservation.

\begin{figure}[t]
    \centering
    \includegraphics[width=0.43\textwidth]{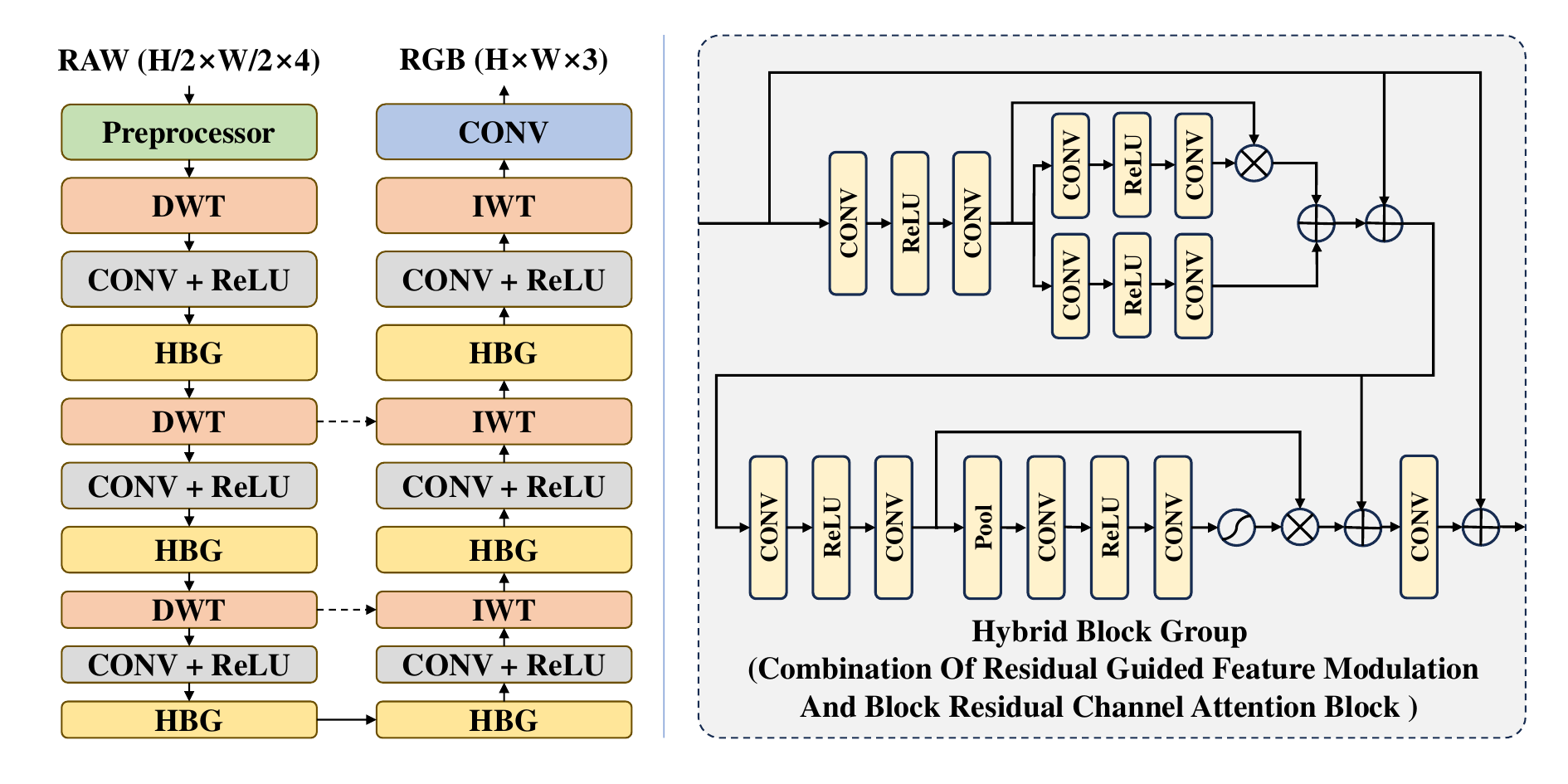}
    \caption{Frame diagram of our method.}
    \label{fig:0005_method}
\end{figure}

\noindent\textbf{Architecture.}  
The model employs a U-shaped encoder-decoder structure enhanced with HybridBlockGroups, which alternate between Residual Guided Feature Modulation (ResGFM) blocks and Residual Channel Attention (RCA) blocks. These modules are designed to balance spatial detail preservation and channel-wise feature enhancement. Additionally, DWTForward and DWTInverse modules are incorporated to perform multi-scale feature extraction and reconstruction using discrete wavelet transforms. The architecture is illustrated in Fig.~\ref{fig:0005_method}.

\noindent\textbf{Training Strategy.}  
The model is trained in two phases using a custom PyTorch framework that supports distributed training and mixed precision for improved efficiency. The first phase focuses on robustness, using heavy data augmentation to train the model against severe low-light noise and artifacts. Augmentations include ISO-based noise modeling (simulating shot and read noise across ISO levels from 3200 to 9600), adaptive Gaussian blur (kernel sizes 3--7 and sigma 0.6--1.2), color channel perturbation (random scaling in [0.8, 1.2]), and dynamic noise amplification (factors between 20 and 30). In the second phase, the model is fine-tuned with lighter augmentations to better preserve texture and fine details while maintaining denoising performance. The dataset used includes 50 images from SIDD~\cite{abdelhamed2018high}, 90 images from SID~\cite{chen2018learning}, and 20 additional images collected from online sources. The training loss is a composite of L1 loss, SSIM loss, and perceptual losses (LPIPS and DISTS), which together balance pixel accuracy, structural similarity, and perceptual fidelity. The entire training process takes approximately 4 hours on a single NVIDIA RTX 3090 GPU. The lightweight design of the network (only 0.83M parameters) allows for efficient training and deployment, requiring minimal human supervision.

\subsubsection{Team xianggkl}
Team xianggkl proposed a diffusion-based framework for joint raw image denoising and demosaicing, named Enhanced Degradation Adaptation Diffusion. The method leverages a trainable VAE encoder $E_{\theta}$, a LoRA fine-tuned two-step diffusion model $\epsilon_{\theta}$, and a frozen VAE decoder $D_{\theta}$. To guide the diffusion process, text prompts extracted from low-quality images are used as conditioning inputs. These prompts help the model adaptively generate high-quality images by aligning the output distribution with that of natural, clean images using Variational Score Distillation (VSD).

\begin{figure}[t]
    \centering
    \includegraphics[width=0.45\textwidth]{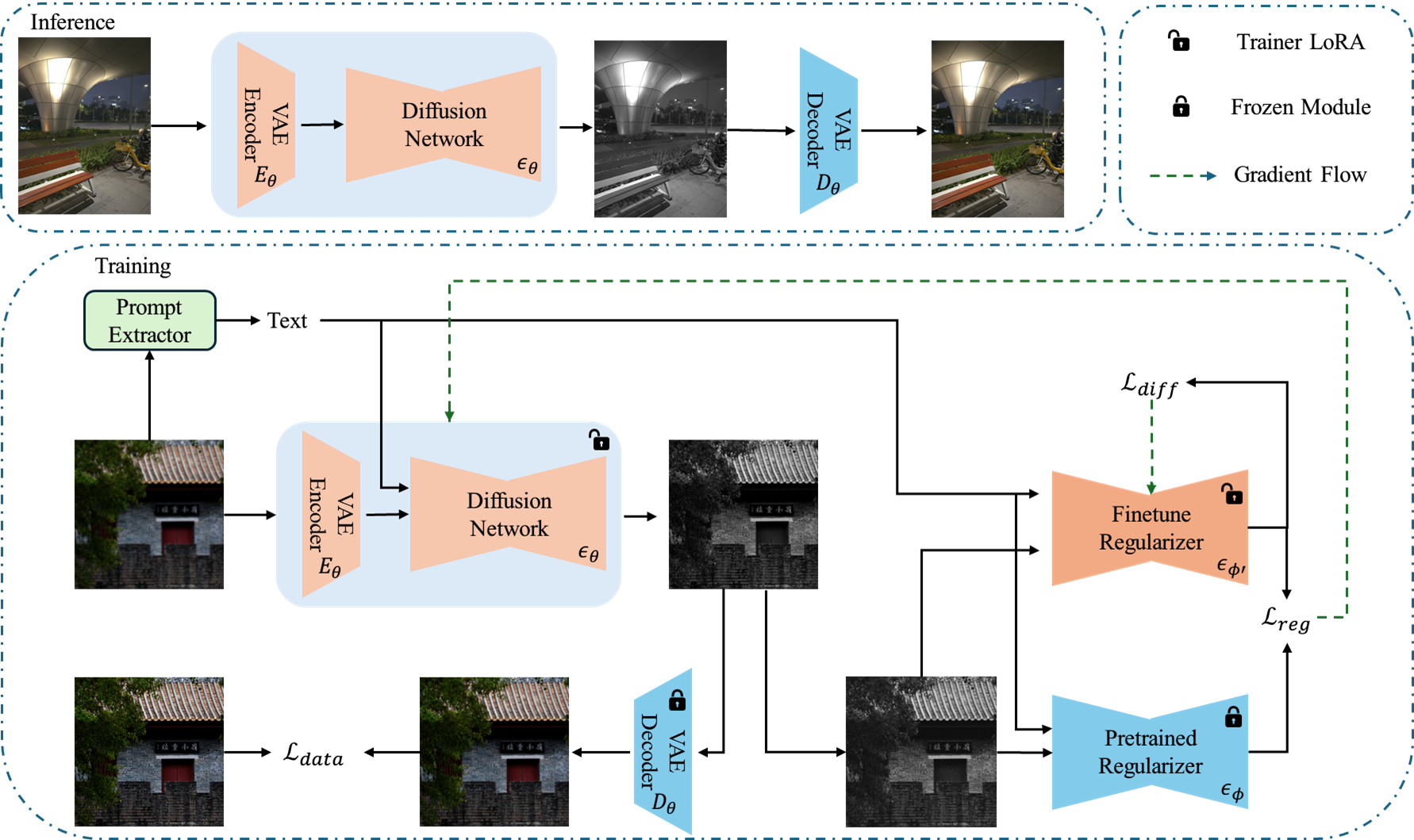}
    \caption{Frame diagram of Team xianggkl.}
    \label{fig:0006_method}
\end{figure}

\noindent\textbf{Architecture and Inference.}  
As shown in the Fig. \ref{fig:0006_method}, the proposed model is based on a diffusion framework enhanced with a trainable VAE encoder $E_{\theta}$, a LoRA fine-tuned two-step diffusion network $\epsilon_{\theta}$, and a frozen VAE decoder $D_{\theta}$. During training, the diffusion output is regularized by two networks—one frozen and one fine-tuned—via VSD in the latent space. The overall objective function combines a data loss $\mathcal{L}_{\text{data}}$ (comprising MSE and LPIPS losses) and a regularization loss $\mathcal{L}_{\text{total}} = \mathcal{L}_{\text{data}} + \lambda_2 \mathcal{L}_{\text{reg}}$,  
 where $\lambda_2$ controls the trade-off between fidelity and distribution alignment. At inference, only $E_{\theta}$, $\epsilon_{\theta}$, and $D_{\theta}$ are used. The prompt extractor and CLIP encoder are removed and replaced with a fixed empty-string embedding. The decoder operates in a fast VAE decoding mode, achieving 37.48 seconds per image of size $3072 \times 4096$.

\noindent\textbf{Training Strategy and Efficiency.}  
Training is conducted in three stages to progressively improve performance and robustness. In Step 1, the model is trained on $256 \times 256$ crops using four RTX 4090 GPUs (batch size 4) for 60K steps. Step 2 continues training with noise-enhanced data (high ISO noise) for 20K steps on two RTX 4090 GPUs (batch size 2). Step 3 fine-tunes the model on $512 \times 512$ crops using two A100 GPUs for 50K steps. Despite the large model size and complexity (2124.36 GFLOPs for $512 \times 512$ input), the system is optimized for deployment with fast inference and minimal runtime overhead.

\noindent\textbf{Novelty and Generalization.}  
The proposed method builds upon OSEDiff~\cite{wu2024one} and incorporates degradation-aware prompt extraction (DAPE) from SeeSR~\cite{wu2024seesr}. Notable innovations include noise-adaptive training, CLIP-free inference via fixed prompts, progressive training stages, and the use of VSD for distribution alignment. The model generalizes well across varying lighting and noise conditions and can be adapted to other restoration and degradation tasks. To the best of the authors’ knowledge, this solution has not been previously published.

\noindent\textbf{Technical Implementation.}  
The system is implemented in Python using PyTorch and trained on 4× RTX 4090 and 2× A100 GPUs. Each experiment takes approximately 2 days. Validation includes both quantitative metrics (PSNR, SSIM, LPIPS) and qualitative visual inspection. Hyperparameters such as latent tile size are tuned through multiple test runs to optimize output quality.

\subsection{Track 2}

\subsubsection{Team NulltoZero}

Team NulltoZero introduces WaveletFusionNet, a deep learning architecture for image detail enhancement, specifically designed for competitive image enhancement tasks. 

\begin{figure}[t]
  \centering
  \includegraphics[width=0.5\textwidth]{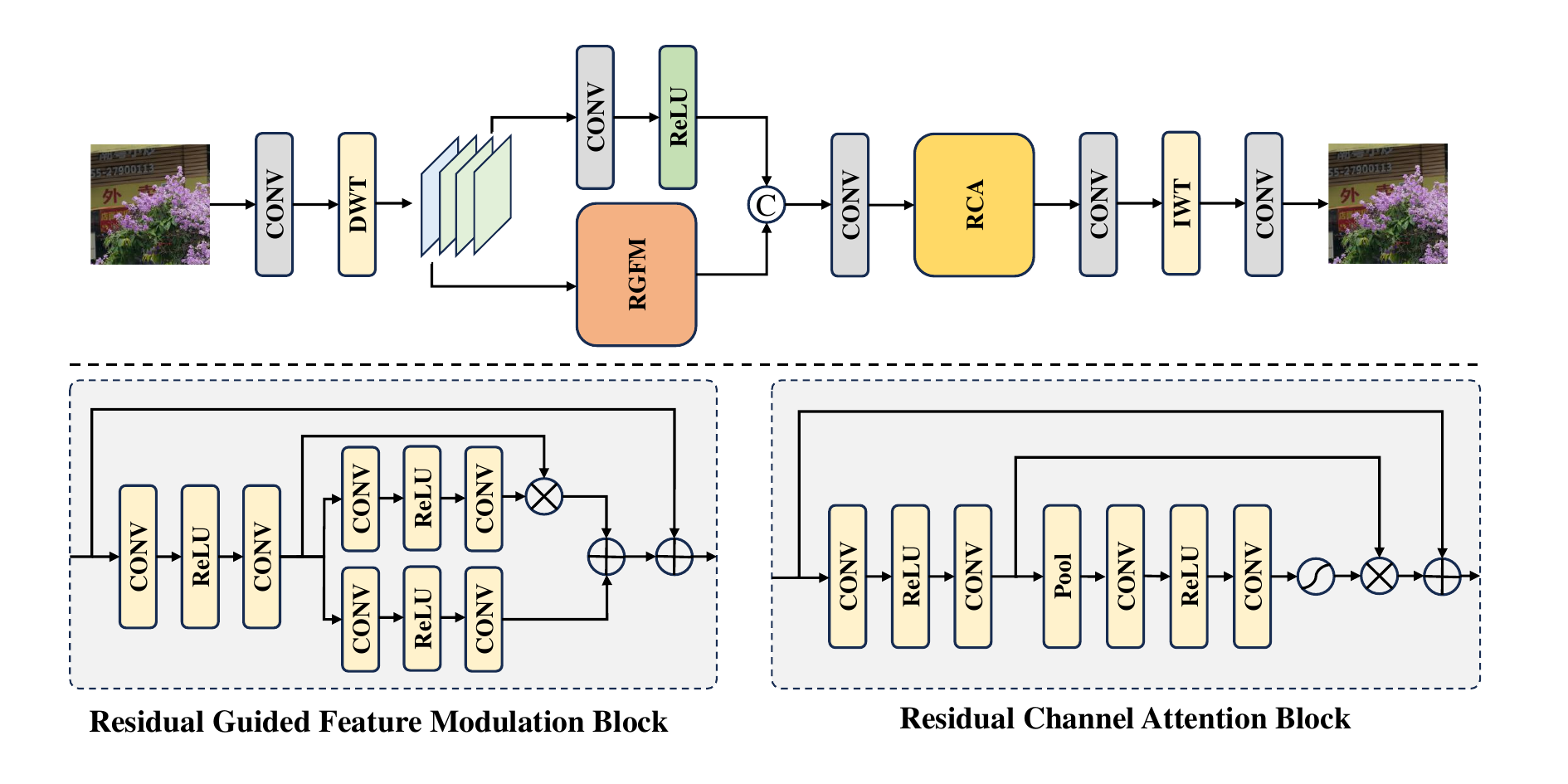} 
  \caption{The overall pipeline of the solution proposed by team NulltoZero.} 
  \label{Figure NulltoZero} 
\end{figure}

\noindent\textbf{Architecture}.As shown in Fig. \ref{Figure NulltoZero}, the proposed model introduces WaveletFusionNet, a novel deep learning architecture tailored for high-resolution image detail enhancement tasks. The model adopts a multi-stage U-shaped framework built upon multilevel discrete wavelet transforms (DWT)to decompose the input into low-frequency and high-frequency components. This decomposition facilitates effective noise suppression and detail recovery.

To extract and enhance features at different scales, the model employs Hybrid Feature Extraction, Multi-Scale Processing, and Fusion and Reconstruction.

\begin{itemize} \item \textbf{Hybrid Feature Extraction Modules}. The network utilizes Residual Channel Attention Groups (RCAGroup) to refine global features and a novel Residual Guided Feature Modulation Block (ResGFMBlock) to adaptively modulate spatial features via a Spatial Feature Transform module.

\item \textbf{Multi-Scale Processing}. The network utilizes Residual Channel Attention Groups (RCAGroup) to refine global features and a novel Residual Guided Feature Modulation Block (ResGFMBlock) to adaptively modulate spatial features via a Spatial Feature Transform module.

\item \textbf{Fusion and Reconstruction}. The enhanced low- and high-frequency features are fused using a $1 \times 1$ convolution followed by further refinement with attention-based modules and residual connections. 

\end{itemize}

The overall design results in an end-to-end system capable of enhancing image details and suppressing artifacts, making it particularly effective for high-resolution image enhancement tasks. The model has approximately 0.303MB of parameters, which have achieved excellent
results in image processing efficiency. The training time of the model can be controlled within 4 hours. When the input size is 1024$\times$1024$\times$3, the reasoning time is about 22.58ms per image on a computer with an RTX3090 GPU. 

\noindent\textbf{Data Augmentation}. Three main steps are involved. First, the target image is read from a specified path, and a corresponding input image path is generated, and the existence of the input image is verified. Then, a random degradation strategy is applied. If the input image exists and a random probability threshold is met, or when not in training mode, the input image is used directly. Otherwise, a random degradation method is applied, either Gaussian blur or bilateral filtering, which is selected randomly. For Gaussian blur, parameters such as kernel size and sigma are randomly sampled, while for bilateral filtering, sigmaColor, sigmaSpace, and the kernel size are chosen randomly. Lastly, synchronized transformation is performed by generating a random seed to ensure that both the degraded input image and target image undergo identical transformations.

\noindent\textbf{Training Details}. The training process using PyTorch starts with loading the training and validation datasets using the DataLoader. The data sets are split based on a set validation ratio. The WaveletFusionNet model is built and trained using the AdamW optimizer. A MultiStepLR scheduler is used to adjust the learning rate during training. The loss function combines $\mathcal{L}_1$ loss, SSIM loss, and a custom NTIRE score loss based on LPIPS and DISTS to support multi-objective optimization. During the validation process, performance is measured using PSNR, SSIM, LPIPS, DISTS, and NIQE. The training loop includes repeated training and validation steps. The checkpoints are saved regularly and the best model is selected on the basis of the validation results. Every five epochs, a visualization is created by randomly selecting an image from the validation set to track the progress of the model.

\subsubsection{Team TeleAI-Vision}

Team TeleAI-Vision adopts the HAT \cite{HAT} architecture to perform the
image restoration task, benefiting from its strong representation capacity and competitive performance.

\noindent\textbf{Detailed Method Description}. In their proposed approach, the team adopts a multi-phase strategy to address the image restoration task. In Phase 2, they utilize the HAT-GAN architecture, leveraging its strong representation capacity and competitive performance to perform high-quality image restoration. To further enhance the model's efficiency, knowledge distillation is applied by using HAT-S as the student network, thereby reducing the model size while preserving performance.

Recognizing the increasing importance of computational efficiency and the demand for deployment-friendly models, the team introduces a more lightweight solution in Phase 3. Specifically, they adopt the Swift Parameter-free Attention Network (SPAN) \cite{span2023} as the backbone of their final model. In their implementation, the number of feature channels is set to 52 and the upscale factor is fixed at 1, in accordance with the specific requirements of the restoration task.

The training process is carried out in three distinct stages. Unlike the Phase 2 model, which relies solely on the provided paired data, the team begins by pretraining the model using the high-quality and diverse LSDIR dataset \cite{li2023lsdir}. This dataset offers a broad range of scenes, enabling the model to acquire a strong and generalizable representation. During the first two stages, the model is trained on the LSDIR training subset with synthetic degradations, allowing it to learn a robust prior. In the third and final stage, the model is fine-tuned using real paired data provided by RAIM 2024 \cite{raim2024} and RAIM 2025. This step is crucial for adapting the model to the target domain and improving its performance under real-world degradation conditions.

Overall, the experimental results demonstrate the effectiveness of the proposed method and underscore the value of carefully curated training data. By progressively leveraging both synthetic datasets and real paired samples, the model achieves strong generalization capabilities while maintaining high adaptability to real-world degradations.

\noindent\textbf{Training Details}. In the phase 2 of their approach, the team trains the HAT model using only the real paired datasets provided by RAIM 2024 \cite{raim2024} and RAIM 2025. To enhance efficiency, knowledge distillation is applied with HAT-S serving as the student network.

In the third phase, with a focus on computational efficiency, the team adopts the Swift Parameter-free Attention Network (SPAN) as the backbone of the model. The complete training process is conducted on 8 GPUs with a batch size of 12. During the first two training stages, the model is trained on the LSDIR dataset \cite{li2023lsdir}, which includes synthetically degraded images. Before training, LSDIR images are cropped into patches of size $240 \times 240$ with a stride of 120. During training, random patches of size $128 \times 128$ are further sampled as model inputs to improve generalization.

The degradation pipeline is based on Real-ESRGAN~\cite{realESRGAN}, with additional degradation hyperparameters to introduce more variability. The degradation settings are as follows:

\begin{itemize}
    \item Gaussian noise probability: 0.5
    \item Noise range: [0.5, 1]
    \item Poisson scale range: [0.05, 0.1]
    \item Gray noise probability: 0.4
    \item Second blur probability: 0.8
    \item Gaussian noise probability (second): 0.5
    \item Noise range (second): [0.5, 1]
    \item Poisson scale range (second): [0.05, 0.1]
    \item Gray noise probability (second): 0.4
\end{itemize}

The model is trained in three stages:

\begin{itemize} \item \textbf{Stage 1} The model is trained using only the $\mathcal{L}_1$ loss on the LSDIR dataset. This stage runs for approximately 330000 iterations with a learning rate of $1 \times 10^{-4}$.
\item \textbf{Stage 2} The Real-SPAN model is trained using a composite loss function to improve perceptual quality and robustness. The loss function is defined as $
\mathcal{L} = \mathcal{L}_{\text{L}_1} + 0.1 \times \mathcal{L}_{\text{Perceptual}} + 4 \times \mathcal{L}_{\text{LPIPS}} + 0.1 \times \mathcal{L}_{\text{GAN}}$.
This stage is trained for approximately 400000 iterations with the same \item \textbf{Stage 3}. The model is fine-tuned using the real paired data from RAIM 2024 and RAIM 2025. Before training, the RAIM images are cropped into patches of size $480 \times 480$ with a stride of 240. The same composite loss function from Stage 2 is used. Fine-tuning is performed for 10000 iterations with a reduced learning rate of $1 \times 10^{-5}$ to ensure stable convergence.
\end{itemize}

\subsubsection{Team TongJi-IPOE}

\begin{figure}[t]
	\centering
	\includegraphics[width=0.43\textwidth]{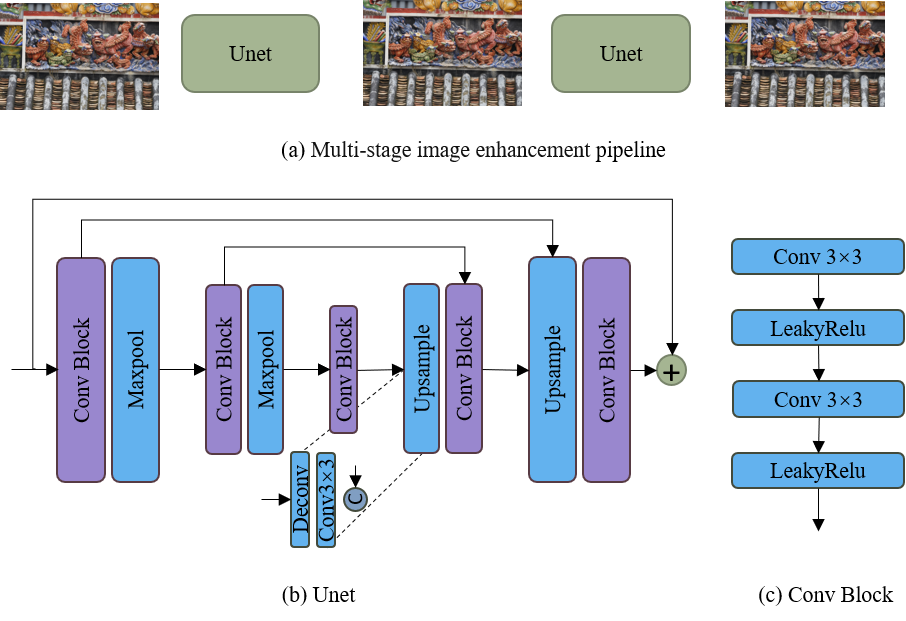}
	\caption{Overview of the MSUnet proposed by Team TongJi-IPOE.}
	\label{Figure TongJi}
\end{figure}

Team TongJi-IPOE proposed a lightweight method for efficient image
detail enhancement/generation on RGB images, named MSUnet (from
branching to fusion network, as shown in Fig. \ref{Figure TongJi}, a multi-stage image restoration network.

\noindent\textbf{Architecture}. Recent research has demonstrated that multi-stage network architectures are highly effective for various image restoration tasks. Compared to directly increasing the number of channels in convolutional layers, cascading multiple networks provides a more lightweight and computationally efficient design~\cite{zamir2021multi}.
Based on this motivation, the team proposes a multi-stage architecture named MSUnet, as illustrated in Fig.~\ref{Figure TongJi} (a). The MSUnet consists of two cascaded U-Net sub-networks. The first U-Net is responsible for performing a preliminary enhancement on the input image, while the second U-Net is designed to further refine the image details and improve the final restoration quality. Both sub-networks adopt a simple yet effective design, consisting solely of $3 \times 3$ convolutional layers, max-pooling layers for downsampling, and deconvolution layers for upsampling.
To achieve better computational efficiency, the team also introduces a lightweight backbone named \textbf{B2FNet}, which serves as the core component of the MSUnet. The B2FNet contains approximately 0.26 M parameters and requires 54.96 GFLOPs to process an image of resolution $1024 \times 1024$. The average inference time of the network is 15 ms on a single NVIDIA A100 GPU, demonstrating its suitability for real-time applications and deployment on resource-constrained platforms.

\noindent\textbf{Training Details}. The entire training process is conducted using a single NVIDIA GeForce RTX 3090Ti GPU, and the implementation is based on the PyTorch framework. The team employs the AdamW optimizer to train the proposed network, with the hyperparameters set to $\beta_1 = 0.9$ and $\beta_2 = 0.999$.
The training is carried out in two stages. In the first stage, the model is trained for 92000 iterations using a fixed learning rate of $1 \times 10^{-4}$. In the second stage, the model undergoes further training for 208000 iterations. During this stage, the learning rate is gradually decreased following a cosine annealing schedule, with a minimum learning rate of $1 \times 10^{-6}$.
Throughout both training stages, only the official dataset provided by the RAIM organizers is used. The training inputs are cropped into patches of size $128 \times 128$ to balance memory usage and training efficiency.

\subsubsection{Team NJUST-KMG}
\noindent\textbf{Method Description}. The team optimizes the MW-ISPNet architecture~\cite{ignatov2020aim} to develop a lightweight model suitable for real-world applications, particularly in resource-constrained environments. Specifically, the number of channels in the downsampling stages is adjusted to 16, 32, and 32, while only a single intermediate layer is retained to simplify the network structure. Correspondingly, the upsampling stages are configured with 32, 32, and 16 channels.
To enhance the model's ability to capture fine-grained image details and textures, several auxiliary loss functions are introduced in addition to the standard $\mathcal{L}_1$ loss. The SSIM loss~\cite{ssim} is employed to preserve structural consistency, with a loss weight of 0.15. Furthermore, LPIPS~\cite{lpips} and DISTS~\cite{dists} losses are incorporated to improve perceptual and texture-level fidelity, both with weights set to 1.0. The combination of these loss functions significantly improves the model’s capacity to restore realistic and detailed image content.

\begin{figure}[h]
  \centering
  \includegraphics[width=1.0\linewidth]{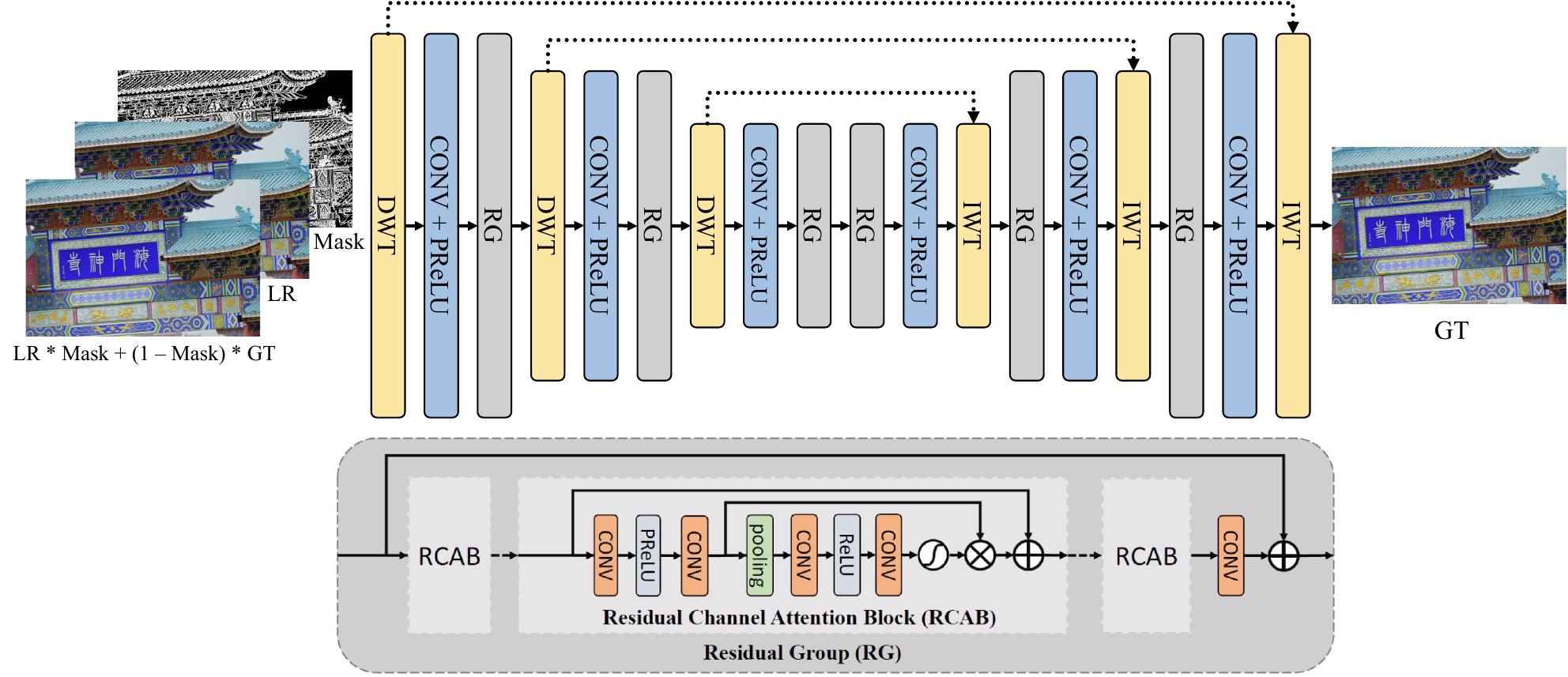}
  \caption{Overall training framework of the method proposed by the team NJUST-KMG.}
  \label{fig NJUST}
\end{figure}

\noindent\textbf{Training and Testing Details}. 
To improve generalization and robustness, data augmentation techniques are applied during training. These include randomly cropping the input images to a resolution of $1024 \times 1024$ pixels, as well as performing random horizontal and vertical flips to diversify the training samples.
In the third phase of development, the team observed that training exclusively with the LR and GT image pairs provided by the organizers often resulted in the generation of severe false textures, which are undesirable in real-world applications. Analysis suggests that this issue arises because of the significant loss of detail and texture information in the LR images, which increases the learning difficulty for the network.
To mitigate this problem, a data pre-processing strategy is proposed to generate LR images that better match the statistical distribution of real mobile phone photographs. As illustrated in Fig.~\ref{fig NJUST}, a slight boundary blur is applied to the GT images. To preserve potential noise in flat regions, the blurring is restricted to edge areas only. The process is as follows: the Canny edge detection algorithm is used to extract an edge mask from the GT image. Then, a Gaussian-blurred version of the GT image is multiplied by the edge mask and added to the original GT image weighted by the inverse mask ($1 - \text{mask}$), resulting in a synthesized LR image that preserves noise in flat areas while reducing sharp transitions at edges. This pre-processing strategy effectively reduces false texture artifacts and enhances the model’s ability to generate high-quality images that align more closely with real-world visual characteristics.

\subsubsection{Team iAM\_IR}

Team iAM\_IR proposed a two-stage image restoration model (TSIRM) to address simulated degradation in phase 2 and real-world degradation in phase 3. In the first stage, the model is pre-trained using extensive simulated data. In the second stage, the model is fine-tuned using GT.

\begin{figure}[t]
\begin{center}
\includegraphics[width=1.\linewidth]{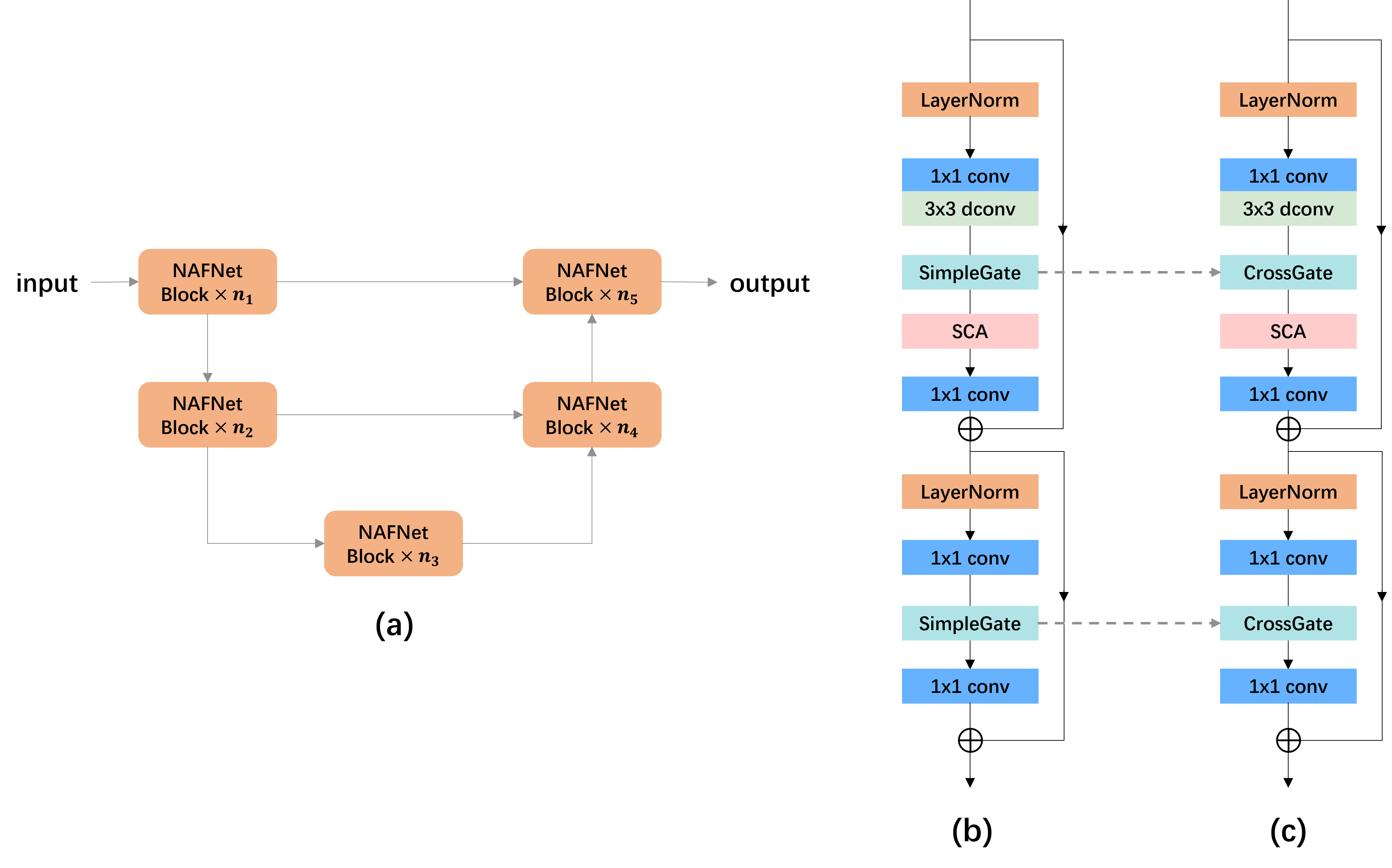}
\end{center}
    \caption{Overall framework proposed by iAM\_IR. (a) represents the U-Net architecture of NAFNet, (b) represents the original NAFNet block, and (c) represents our improved NAFNet block.}
\label{fig zhenyu-arch}
\end{figure}

\begin{figure}[t]
    \centering
    \includegraphics[width=1.\linewidth]{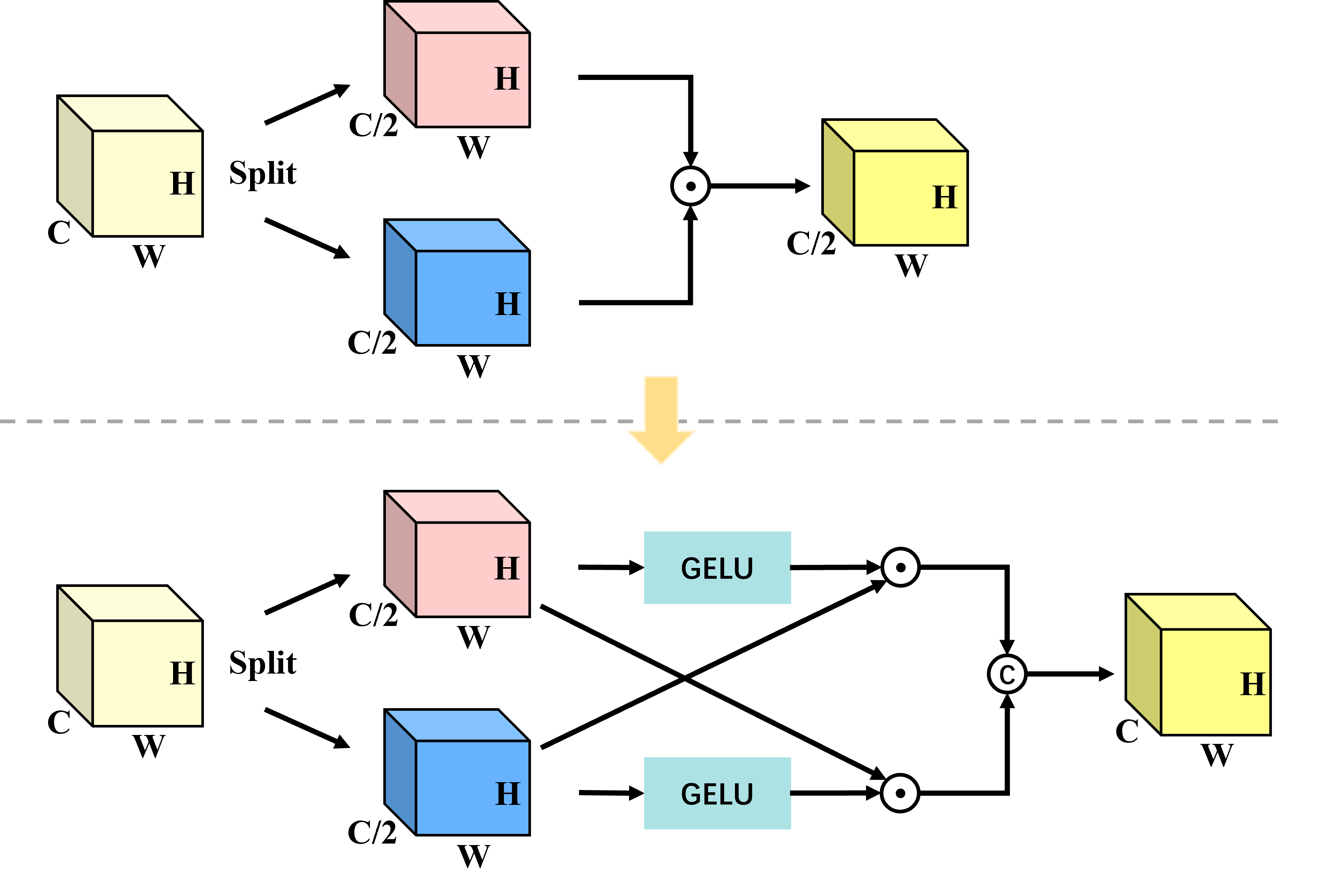}
    \caption{Structure comparison between SimpleGate and the CrossGate proposed by iAM\_IR.}
    \label{fig zhenyu-crossgate}
\end{figure}

\noindent\textbf{Architecture}.
Considering the efficiency and practical deployment requirements, the proposed model is built upon NAFNet~\cite{NAFNet}, with several modifications to improve feature representation and model performance. As illustrated in Fig.~\ref{fig zhenyu-arch}, the model adopts a U-Net-like architecture composed of multi-scale NAFNet blocks.
In the original NAFNet design, the GELU activation function is replaced with the computationally efficient SimpleGate operation, and the self-attention mechanism is substituted with a lightweight channel attention module. These changes significantly improve computational efficiency while preserving the benefits of transformer-based architectures.
However, SimpleGate alone may result in limited feature interaction. To address this limitation, the team introduces an improved activation mechanism named CrossGate, as shown in Fig.~\ref{fig zhenyu-crossgate}. Unlike SimpleGate, the CrossGate module enables more effective fusion of complementary feature information without reducing channel dimensionality, thereby enhancing the model's representation capacity.

\noindent\textbf{Two-Stage Training Strategy}.
Due to the limited availability of real paired training data, a two-stage training strategy is adopted. 
In the first stage, the model is pretrained using synthetic paired data generated from high-quality public datasets. Specifically, the team uses the DIV8K~\cite{div8k} dataset and 1,000 facial images from FFHQ \cite{ffhq}. To simulate realistic degradations, a simplified version of the BSRGAN~\cite{BSRGAN} pipeline is adopted, in which JPEG compression is removed to reduce artifacts. The commonly used Real-ESRGAN~\cite{realESRGAN} second-order degradation pipeline is avoided due to its tendency to introduce overly severe distortions.
Pretraining is conducted in two steps: first, 20,000 iterations are trained using only $\mathcal{L}_1$ loss; then, perceptual loss and GAN loss are introduced, with the loss function defined as:
\[
\mathcal{L}_{\text{total}} = \mathcal{L}_{1} + 1.0 \times \mathcal{L}_{\text{perceptual}} + 0.1 \times \mathcal{L}_{\text{GAN}},
\]
and training continues for an additional 150000 iterations.
In the second stage, fine-tuning is performed using available high-quality reference images. In Phase 2, where GT data is available, the team employs all reference-based evaluation metrics from the competition as supervisory objectives to directly optimize image quality and perceptual fidelity.
In Phase 3, where no GT is provided, pseudo GT images are generated using FeMaSR~\cite{chen2022femasr} (×2 version), a state-of-the-art real image restoration model. The generated pairs are then used to fine-tune the network. To further enhance perceptual quality, the weights of the perceptual and GAN losses are increased to:
\[
\mathcal{L}_{\text{total}} = \mathcal{L}_{1} + 1.5 \times \mathcal{L}_{\text{perceptual}} + 0.2 \times \mathcal{L}_{\text{GAN}}.
\]

\subsubsection{Team MiAlgo}

\noindent\textbf{Phase2}.
The MiAlgo team fine-tuned the model for approximately 20000 iterations using the loss function $\mathcal{L}_{2} + 0.1 \cdot \mathcal{L}_{\text{perceptual}} + 0.01 \cdot \mathcal{L}_{\text{GAN}} + 4 \cdot \mathcal{L}_{\text{lpips}}$. The fine-tuning process was conducted within the RealESRGAN \cite{realESRGAN} training framework, employing a learning rate of 1e-5. For training data, the degraded data from the RAIM2024 \cite{raim2024} challenge was combined with the 50 paired data samples provided by RAIM2025, in a 50:50 ratio in the training filelist.

\begin{figure}[t]
    \centering
    \includegraphics[width=1.\linewidth]{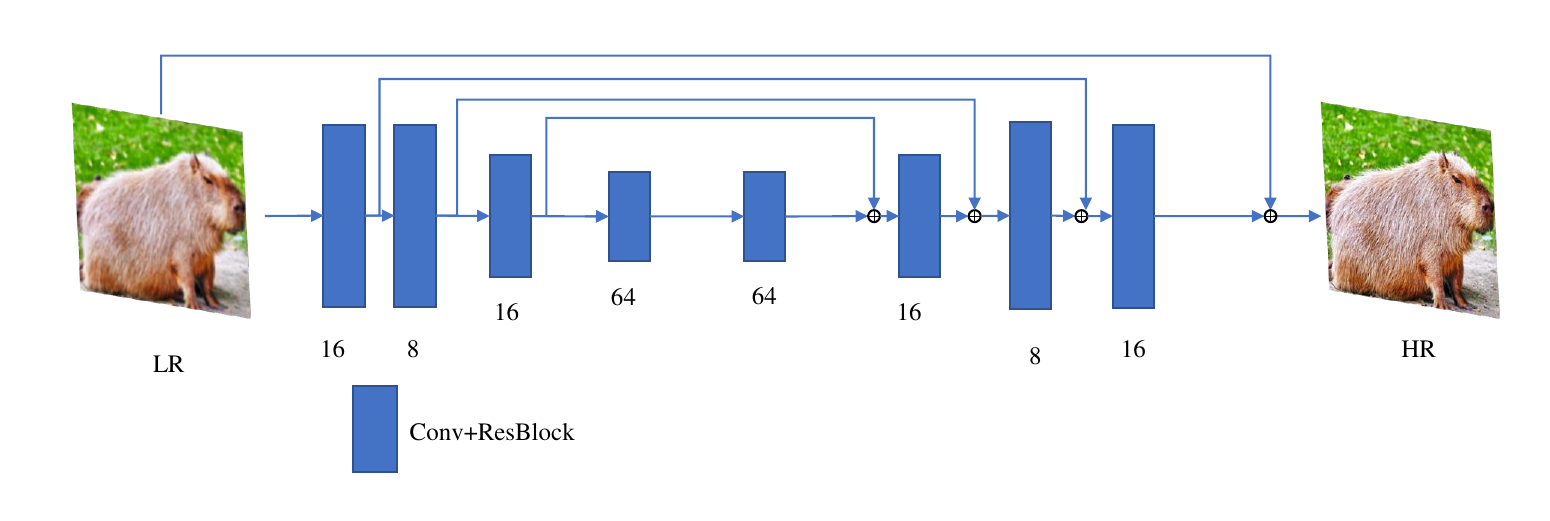}
    \caption{The tiny unet in phase3 applied by the team MiAlgo.}
    \label{fig MiAlgo_Unet-tiny}
\end{figure}

\noindent\textbf{Phase3}.
As shown in Figure \ref{fig MiAlgo_Unet-tiny}, the Phase 3 model is built upon the Phase 2 architecture, adopting a lightweight UNet structure enhanced with Haar wavelet-based downsampling and upsampling (×2). The design is inspired by the MWRCAN model \cite{ignatov2020aim}, where each UNet block contains one ResBlock, with convolutional layers between ResBlocks and wavelet modules to adjust channel dimensions. Given the high quality of Phase 3 test data, only minor edits were necessary. To maintain fidelity to the input, a global residual connection was added into the network design to ensure the output remains closely aligned with the original image.
This efficient configuration results in a model size of only 0.075 MB and 19.83 GFLOPs for $3 \times 1024 \times 1024$ inputs. On an A100 GPU, it processes this input in 5.95 ms and handles $3 \times 3040 \times 4032$ images in 58.90 ms.

The training dataset consisted of two parts: one inherited from Phase 2, and another newly synthesized. The latter was motivated by the observation that 50\% of the test images were high-quality female portraits. Approximately 5000 curated DSLR portrait of young women were curated and used as ground truth. These images were converted to RAW format and augmented with noise. 
Subsequently, the Track 1 Phase 2 model was employed to render these images into RGB format, serving as the low-quality images. Similarly, the GT used in phase 2 was also processed in the same manner to generate low-quality images.

Training began from scratch using the full dataset. The model was first optimized with $\mathcal{L}_2$ loss for 10k steps, followed by 800k steps using a composite loss: $\mathcal{L}_2 + 0.1 \cdot \mathcal{L}_{\text{perceptual}} + 0.01 \cdot \mathcal{L}_{\text{GAN}}$. A learning rate of $1e^{-5}$, batch size of 32, and input size of 512 were employed across 4 GPUs.
To emphasize portrait quality, the dataset was rebalanced to include 80\% portrait images. A learning rate of $1e^{-6}$ was used for an additional 40k steps. In the final stage, another 10k steps were trained with USM-enhanced ground truths and a reduced learning rate of $1e^{-7}$.
Despite some training data being blurrier than the test set, such degradation was found to improve the clarity of non-portrait scenes.

{
    \small
    \bibliographystyle{ieeenat_fullname}
    \bibliography{main}
}


\end{document}